\documentclass[pra,onecolumn,showpacs,superscriptaddress]{revtex4}
\usepackage{amsmath}
\usepackage{latexsym}
\usepackage{graphicx}
\usepackage{color}
\usepackage{bm}
\usepackage{float}
\usepackage{bbold}
\usepackage{amsfonts,amssymb}
\usepackage[normalem]{ ulem }
\usepackage{soul}
\usepackage{verbatim}
\usepackage{booktabs}
\newcommand{\be}{\begin{equation}}
\newcommand{\ee}{\end{equation}}
\newcommand{\bea}{\begin{eqnarray}}
\newcommand{\eea}{\end{eqnarray}}
\newcommand{\QNS}{\chi_{{\rm Q}N{\rm S}}}

\newenvironment{psmallmatrix}
{\left(\begin{smallmatrix}}
	{\end{smallmatrix}\right)}

\definecolor{blue-violet}{rgb}{0.54, 0.17, 0.89}
\definecolor{chromeyellow}{rgb}{1.0, 0.65, 0.0}
\definecolor{caribbeangreen}{rgb}{0.0, 0.8, 0.6}
\definecolor{light-blue}{rgb}{0.6,0.6,1}
\definecolor{cadmiumgreen}{rgb}{0.0, 0.42, 0.24}
\definecolor{Blue}{rgb}{0,0,1}
\definecolor{Red}{rgb}{1,0,0}

\newcommand{\N}[1]{{\mathcal N}_{#1}}
\DeclareMathOperator{\tr}{Tr}
\usepackage{hyperref} 
\usepackage[all]{hypcap}
\hypersetup{pdfauthor = {First Author},pdftitle = {Communication enhancement through quantum coherent control of $N$ channels in an indefinite causal-order scenario}}

\begin{document}
	
	\title{Communication Enhancement Through Quantum Coherent Control of $N$ Channels in an Indefinite Causal-order Scenario}
	\author{Lorenzo M. Procopio}
	
	\thanks{Corresponding author: lorenzo.procopio@c2n.upsaclay.fr}
	\affiliation{Centre for Nanoscience and Nanotechnology, C2N, CNRS, Universit\'e Paris-Sud, Universit\'e Paris-Saclay, 10 Boulevard Thomas Gobert, 91120 Palaiseau, France}
	
	\author{Francisco Delgado}
	\affiliation{Tecnologico de Monterrey, Escuela de Ingeniería y Ciencias, Carr. a Lago de Guadalupe km. 3.5, Atizap\'an, Estado de M\'exico, M\'exico, 52926}
	
	\author{Marco Enr\'iquez}
	\affiliation{Tecnologico de Monterrey, Escuela de Ingeniería y Ciencias, Carr. a Lago de Guadalupe km. 3.5, Atizap\'an, Estado de M\'exico, M\'exico, 52926}
	
	\author{Nadia Belabas}
	\affiliation{Centre for Nanoscience and Nanotechnology, C2N, CNRS, Universit\'e Paris-Sud, Universit\'e Paris-Saclay, 10 Boulevard Thomas Gobert, 91120 Palaiseau, France}
	
	\author{Juan Ariel Levenson}
	\affiliation{Centre for Nanoscience and Nanotechnology, C2N, CNRS, Universit\'e Paris-Sud, Universit\'e Paris-Saclay, 10 Boulevard Thomas Gobert, 91120 Palaiseau, France}

	\begin{abstract}
	In quantum Shannon theory, transmission of information is enhanced by quantum features. Up to very recently, the trajectories of transmission remained fully classical. Recently, a new paradigm was proposed by playing quantum tricks on two completely depolarizing quantum channels i.e. using coherent control in space or time of the two quantum channels. We extend here this control to the transmission of information through a network of an arbitrary number $N$ of channels with arbitrary individual capacity i.e. information preservation characteristics in the case of indefinite causal order. 
	We propose a formalism to assess information transmission in the most general case of $N$ channels in an indefinite causal order scenario yielding the output of such transmission. Then we explicitly derive the quantum switch output and the associated Holevo limit of the  information transmission for $N=2$, $N=3$ as a function of all involved parameters.  We find in the case $N=3$ that the transmission of information for three channels  is twice  of transmission of the two channel case when a full superposition of all possible causal orders is used.
	\end{abstract}
	\maketitle


\section{Introduction}
In information theory, the main tasks to perform are the transmission, codification, and compression of information \cite{shannon1948mathematical}. Incorporating quantum phenomena, such as quantum superposition and quantum entanglement, into classical information theory gives rise to a new paradigm known as quantum Shannon theory \cite{nielsen2002quantum}.  In this paradigm, each figure of merit can be enhanced: the capacity to transmit information in a channel is increased \cite{holevo1998capacity}, the security to share a message is improved \cite{bennett2014quantum} and the storing and compressing of information is optimized \cite{schumacher1995quantum}. In all these enhancements, only the carriers and the channels of information  are considered as quantum entities. On the other hand,  connections between  channels are still classical, that is,  quantum channels are connected setting a definite causal order in space or time. However, principles of quantum mechanics and specifically the quantum superposition principle can be applied to the connections of channels \cite{Chiribella2013}, i.e. the trajectories either in space \cite{abbott2018communication} or time \cite{chiribella2018indefinite}. 

Recently, it has been theoretically \cite{ebler2018enhanced} and experimentally  \cite{goswami2018communicating,guo2018experimental} shown that two completely depolarizing channels can surprisingly transmit classical information when combined under an indefinite causal order (i.e., when the order of application of the two channels is not one after another instead of a quantum superposition of the two possibilities). In this paper, we tackle the general situation of an arbitrary number $N$ of channels with arbitrary parameters associated to the control and depolarizing strength. As $N$ is greater than two, the number of different causal orders increases as $N!$   

The indefiniteness of causal order has been recently theoretically proposed as a novel resource for applications to quantum information theory \cite{chiribella2012perfect,Araujo2014} and quantum communication \cite{salek2018quantum,guerin2016exponential}. 
Initially, indefinite causal orders have been studied and implemented using two parties with the proposal of a quantum switch by Chiribella et al. \cite{Chiribella2013} followed by experimental demonstrations 
\cite{guo2018experimental,procopio2015experimental, goswami2018indefinite,wei2018experimental,rubino2017experimental}
The quantum switch is an example of quantum control where a switch can, like its classical counterpart, routes a target system to undergo through two operators in series following one causal order ($1$ then $2$) or the other ($2$ then $1$). But this quantum switch can also trigger a whole new quantum trajectory where the ordering of the two operators is indefinite.  Efforts to describe the quantum switch in a multipartite scenario of more than two quantum operations have recently started  \cite{wechs2018definition,oreshkov2016causal} with an application to reduce the number of queries for quantum computation\cite{Araujo2014}. 
\begin{figure} [h!]
	\vspace*{13pt}
	\scalebox{.35}{\includegraphics{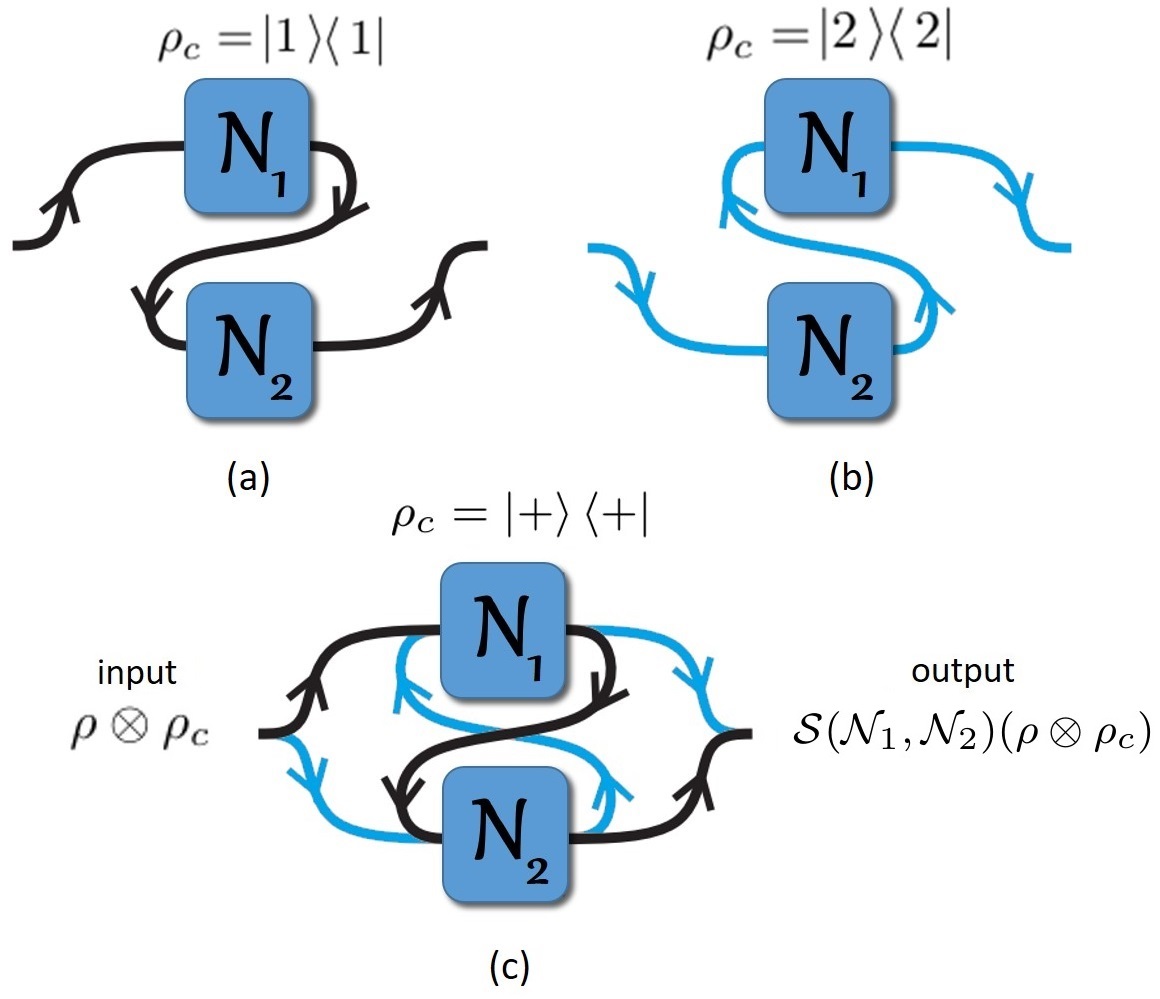}}
	\vspace*{13pt}
	\caption{\label{Figura1} {Concept of the quantum 2-switch.} $\N{i}={\mathcal N}_{q_i}^D$ is a depolarizing channel applied to the quantum state $\rho$, where $1-q_i$ is the strength of the depolarization. For two channels, depending on the control system $\rho_c$, there are 2! possibilities to combine the channels with definite causal order: (a)  if $\rho_c$ is in the state $\left| 1 \right>\left< 1 \right|$, the causal order will be $\N{2}\circ\N{1}$, i.e. $\N{1}$ is before $\N{2}$; (b) on the other hand,  if $\rho_c$ is on the state $\left| 2 \right>\left< 2 \right|$, the causal order will be $\N{1}\circ\N{2}$; (c) however, placing  $\rho_c$ in a  superposition of its states (i.e.  $\rho_c=\left| + \right>\left< + \right|$, where $\left| + \right>_c= \frac{1}{\sqrt{2}}(\left| 1 \right> + \left| 2 \right> )$) results in the indefinite causal order of  $\N{1}$ and  $\N{2}$ to become indefinite. In this situation we said that the quantum channels are in a superposition of causal orders. This device is called a quantum 2-switch \cite{Chiribella2013} whose input and output are $\rho\otimes\rho_c$ and  ${\mathcal S}(\N{1},\N{2})(\rho\otimes\rho_c)$ respectively. } 
\end{figure}

\begin{figure*} [h!]
	\vspace*{13pt}
	\scalebox{.3}{\includegraphics{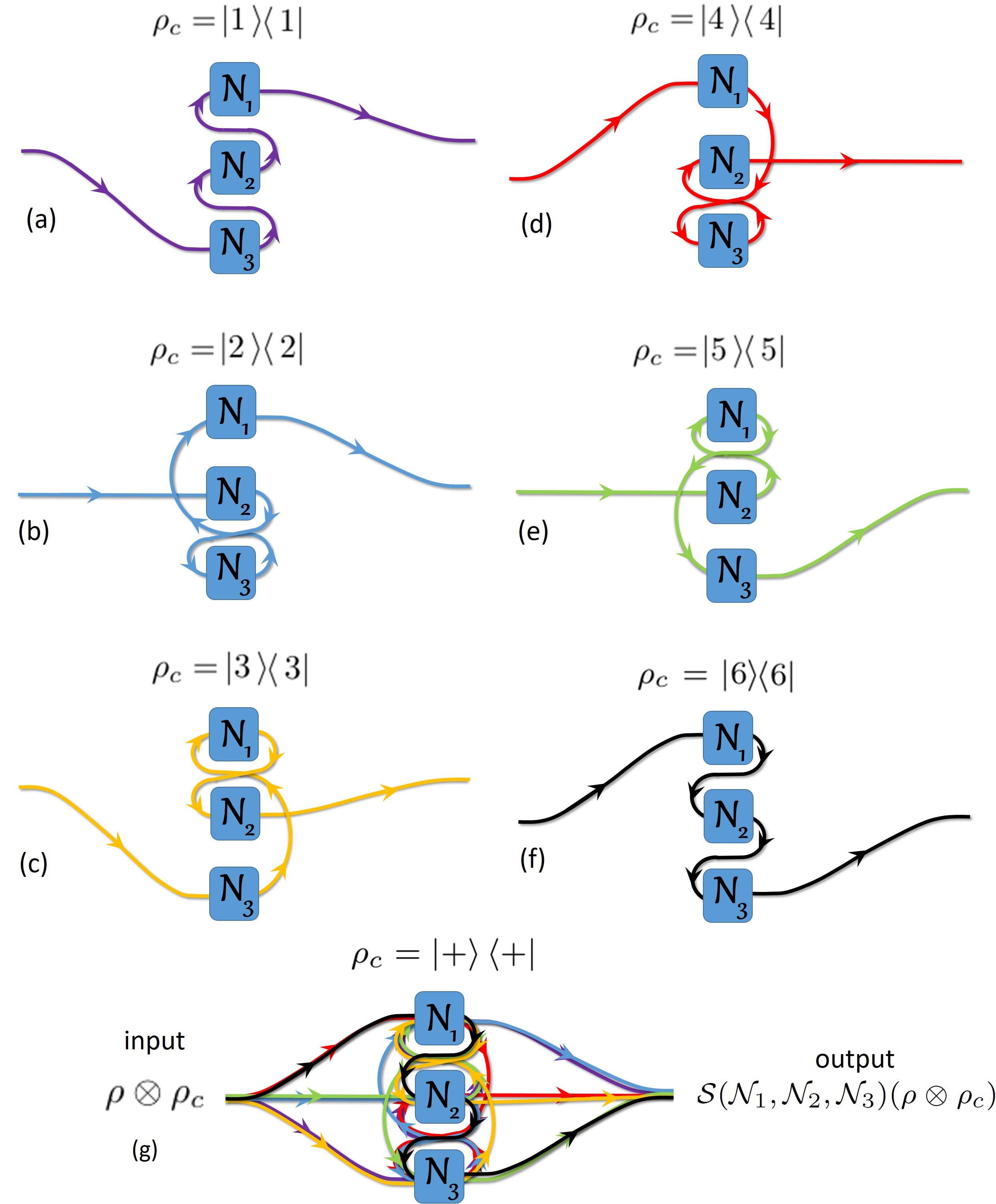}}
	\vspace*{13pt}
	\caption{\label{Figura2} \footnotesize Concept of the quantum 3-switch. For three channels, depending on $\rho_c$, we have 3! possibilities to combine the channels in a definite causal order: 
		(a) $\rho_c=\left| 1 \right>\left< 1 \right|$ encodes a causal order  $\N{1}\circ\N{2}\circ\N{3}$, i.e. $\N{3}$ is applied first to $\rho$; 
		(b) $\rho_c = \left| 2 \right>\left< 2 \right|$ encodes $\N{1}\circ\N{3}\circ\N{2}$;
		(c) $\rho_c = \left| 3 \right>\left< 3 \right|$ encodes $\N{2}\circ\N{1}\circ\N{3}$; 
		(d)  $\rho_c = \left| 4 \right>\left< 4 \right|$ encodes $\N{2}\circ\N{3}\circ\N{1}$; 
		(e)  $\rho_c = \left| 5 \right>\left< 5 \right|$ encodes $\N{3}\circ\N{1}\circ\N{2}$; 
		(f)  $\rho_c = \left| 6 \right>\left< 6 \right|$ encodes $\N{3}\circ\N{2}\circ\N{1}$;
		(g) finally, if $\rho_c=\left| + \right>\left< + \right|$,  where $\left| + \right>= \frac{1}{\sqrt{6}}\sum_{k=1}^6 \left| k \right>$ we shall have a superposition of six different causal orders. This is an indefinite causal order called quantum 3-switch whose input and output are $\rho\otimes\rho_c$ and  ${\mathcal S}(\N{1},\N{2},\N{3})(\rho\otimes\rho_c)$ respectively. Notice that for each superposition with $m$ different causal orders, there are $\binom{N!}{m}$ (with $m=1,2,\ldots,6$ ) possible combinations of causal orders to build such superposition with $N=3$ channels, where $\binom{n}{r}= \frac{n!}{r!(n-r)!}$ is the binomial coefficient. The input and output of each channel are fixed. The arrows along the wire just indicate that the target system enters in or exits from the channel.} 
\end{figure*}

Specifically, in a quantum $N$-switch used in a second-quantized Shannon theory context \cite{chiribella2018second}, the order of application of $N$ channels $\N{j}$ on a target system $\rho$ is coherently controlled by a control system  $\rho_c$. The state of $\rho_c$ encodes for the temporal combination of the $N$ channels applied to $\rho$. There are $N!$ different possibilities of definite causal orders using each channel once and only once, as sketched in Figures \ref{Figura1} and \ref{Figura2} for $N=2$ and $N=3$ respectively. In those figures, when the wiring passes through the channel, there is a single channel use, i.e.  the target system passes once  through one physical channel \cite{procopio2015experimental}. We discard all wirings with multiple use of the same channel and missing channels \cite{abbott2018communication}.  For each causal order of channels, the overall operator is  
\begin{equation}
	\N{\pi} := \pi(\N{1}\circ\dots\circ\N{N})
	\label{CausalProduct}
\end{equation}
where $\pi$ is a permutation element of the symmetric group $ S_N=\{ \pi_k | k \in \{1,2,\ldots,N!\} \}$, and $k$ is associated to a specific  definite causal order (equivalent to a single element of $S_N$) to combine the $N$ channels where each channel is used once and only once.

In a quantum $N$-switch, the control state $\rho_c$ in the state $\left| 1 \right>\left< 1 \right|$ for instance fixes the order of application of the channels to be $\N{\text{Id}}=\N{1} \circ\N{2}\circ\cdots\circ\N{N}$. Whereas, choosing $\rho_c=\left| k \right>\left< k \right|$, $k\leq N!$ would assign another ordering $\N{\pi_k}=\N{\pi_k(1)}\circ\N{\pi_k(2)}\circ\cdots\circ\N{\pi_k(N)}$ (defined by the effect of the permutation element $\pi_k\in S_N$ on the order of channels). 
The key to accessing indefinite causal order of the channels is thus to put $\rho_c$ in a superposition of the $\left| k \right>\left< k \right|$ states (e.g. $\rho_c=\left| + \right>\left< + \right|$ where $\left| + \right>\equiv \frac{1}{\sqrt{N!}}\sum\left| k \right>$).

The paper is organized as follows. Section \ref{framework} is devoted to the general theoretical framework for the investigation of the transmission of classical information over $N$ noisy channels with arbitrary degree of depolarization, i.e. arbitrary level of noise. Section~\ref{framework} also gives the channels representation in terms of Kraus operators performed from those operators for a single depolarizing channel. In Section \ref{results}, following the previous formalism, we explicitly analyze the case $N=2$, generalizing the outcomes in the literature \cite{ebler2018enhanced} to any degree of depolarization and level of coherent control. Similarly, the case $N=3$ is developed in the same section.  Finally, conclusions and perspectives are given in Section \ref{concl}.

\section{Transmission over multiple channels in quantum superposition of causal order }
\label{framework}
\noindent 
In the current development, the sender prepares the target system in the state $\rho$, where the information to transmit is encoded. A control system $\rho_c$ is associated to the target system to coherently control the causal order for the application of $N$ quantum communication channels. We relate the basis for the quantum state $\rho_c$ mapping their elements on those of the symmetric group of permutations $S_N: \frac1{N!} \sum_{k,k'} \vert k\rangle\langle k'\vert$.
Then, the sender introduces as input $\rho\otimes\rho_c$ to a network of $N$ partially depolarizing channels $\N{i}=\mathcal N_{q_i}^D,1\leq i\leq N$ applied in series (i.e. the output of one channel becomes the input of the next channel).
Throughout this work the $N$ depolarizing channels $\N{1}$, $\N{2}, \ldots$, $\N{N}$ can have different depolarization strengths $1-q_j$, (thus, ${\mathcal N}_{q_j}^D$ is sometimes used for $\N{j}$ to improve the readability).

After the network, the receiver gets the output state ${\mathcal S}({\mathcal N}_1,{\mathcal N}_2,\ldots,{\mathcal N}_{N})(\rho\otimes\rho_c)$, where ${\cal S}$ is the quantum $N$-switch channel. No information is encoded by the sender into the control system controlling the way information is transmitted. Eventually, the receiver retrieves the information decoded in  $\rho$. 

Communication quantum channels in a network are mathematically described with completely positive trace preserving maps (CPTP). Here, we adopt the Kraus decomposition \cite{nielsen2002quantum} $\N(\rho)=\sum_{i} K_{i}  \rho K_{i}^{\dagger}$ to describe the action of a total depolarizing channel ${\mathcal N}$ on the quantum state $\rho$ ($i \in\{1,2,\ldots,d\}$): $\N{\rho} = {\tr} [\rho] \frac{\mathbb{I}}{d}$. 
The set of $d^2$ non-unique and generally non-unitary Kraus operators $\{K_{i}\}$ satisfies the completeness condition $\sum_{i=1}^{d^2} K_{i} K_{i}^{\dagger}=\mathbb{I}$. 
Thus, to describe the action of the $j$-th  partially depolarizing channel $\N{j}$ on a $d$-dimensional quantum system $\rho$, we write as in \cite{ebler2018enhanced}
\begin{align} \label{depch2}
	\N{q_j}^D(\rho) &= q_j \rho + (1-q_j) {\tr} [\rho] \frac{\mathbb{I}_t}{d}= q_j\rho + \frac{1-q_j}{d^2} \sum_{i_j=1}^{d^2} U_{i_j}^{j}
	\rho U_{i_j}^{j\dagger}\nonumber \\
	& = \frac{1-q_j}{d^2}\sum_{i_j=0}^{d^2} U_{i_j}^{j} \rho U_{i_j}^{j\dagger}
\end{align}

\noindent where each $\N{j}=\N{q_j}^D$ is thus decomposed on an orthonormal basis $\{U_{i_j}^j\}|_{i_j=1}^{d^2}$. Then, we define $K_{i_j}^{j} = \frac{\sqrt{1-q_j}}{d} U_{i_j}^{j}$ for $i_j \ne 0$, where the added non-unitary operator $U_0^j= \frac{d \sqrt{q_j}}{\sqrt{1-q_j}} \mathbb{I}_t$, for $i_j=0$. Besides $\N{j}$ has no noise when $q_j=1$. On the other hand, $\N{j}$ is completely depolarizing when $q_j=0$. The results reported in \cite{ebler2018enhanced,abbott2018communication} are mainly related to two completely depolarizing ($q_1=q_2=0$) channels $\N{1}$ and $\N{2}$, despite the generalization is outlined. Below we extend the results from \cite{ebler2018enhanced} to the case of a quantum switch with $N$ channels ${\mathcal N}_{j}$  with arbitrary individual depolarization strengths $q_j$.

\subsection{The formalism for a quantum $N$-switch channel ${\mathcal S}(\N{1},\N{2},\ldots,\N{N})$ }

We define the control state $\rho_c$ as $\rho_c=\left| \psi_c \right>\left< \psi_c \right| = \sum_{k,k'=1}^{N!} \sqrt{P_k P_{k'}} \left| k \right>\left< k' \right|$ where $P_k$ is the  probability to apply the causal order $k$ (corresponding to the permutation $\pi_k$ as it was previously stated) to the channels such that $\sum_{k=1}^{N!} P_k=1$.

The action of the quantum $N$-switch channel ${\mathcal S}(\N{1},\N{2},\ldots,\N{N})$ can be expressed through generalized Kraus operators $W_{i_1 i_2\ldots i_N}$ for the full quantum channel resulting from the switching of $N$ channels as
\begin{equation}\label{Krausg}
	\mathcal S(\N{1},\N{2},\ldots,\N{N}) \left( \rho \otimes \rho_c \right)  =  \sum_{\{i_j\}|_{j=1}^N} W_{\mathbf{i}}\left( \rho \otimes \rho_c \right.) W_{\mathbf{i}}^\dagger
\end{equation}

\noindent where $ W_{\mathbf{i}} := W_{i_1 i_2\ldots i_N}=  \sum_{k=1}^{N!} K_{\pi_k} \otimes \left| k \right>\left< k \right|$ and $K_{\pi_k}$ has been defined similarly to equation~(\ref{CausalProduct}) : $ K_{\pi_k} := \pi_k( K_{i_1}^{1}\cdots K_{i_N}^{N})$ where $\pi_k$ acts on the index $j$, and the sum over $\{i_j\}|_{j=1}^N$ means all $i_j$ associated to each channel $\N{j}$ vary from $0$ to $d^2$.  We verify (see Appendix~\ref{AppA}) that these generalized Kraus operators satisfy the completeness property $\sum_{\{i_j\}|_{j=1}^N}  W_{\mathbf{i}}  W_{\mathbf{i}}^\dagger= \mathbb{I}_t \otimes \mathbb{I}_c $, where identity operators in the target  and  control systems spaces are denoted $\mathbb{I}_t$ and $\mathbb{I}_c$, respectively. This check of completeness suggests how the $i_j$ indices allow the systematic reordering of the sums by isolating and grouping the $i_j=0$ cases. 
To distinguish those terms, we introduce the number $z$ of indices $i_j$ equal to zero. The sums over the indices $i_j$ can then be rearranged as 
\begin{equation}\label{sumABz}
	\sum_{\{i_j\}|_{j=1}^N} \rightarrow \sum_{z=0}^N \hspace{1ex}\sum_{A_z\in {\bf A}_z^N}  \sum_{b\in B_z},
\end{equation}
where $A_z$ is the set of $z$ indices equal to zero ($i_a=0$, $\forall a \in A_z$) and $B_z$ is the complementary set of indices in $\{1,2\ldots,N\}$ : $i_b \neq 0$, $i_b\in \{1,2,\ldots,d^2\}$ for all $b \in B_z$. Then, $U_{\pi_k} U_{\pi_k}^\dagger=d^{2z} h_{A_z}\mathbb{I}_t$, where $h_{A_z}=\prod_{a\in A_z} \frac{q_a}{1-q_a}$ and $h_{A_0}=1$.

Introducing the Kraus operators $W_{\mathbf{i}}$ into $\mathcal S(\N{1},\N{2},\ldots,\N{N})$, equation (\ref{Krausg}) can be written as a sum of $N+1$ matrices ${\cal S}_z$ whose $N! \times N!$ elements are matrices of dimension $d \times d $ involving exactly $z$ factors $U_{i_j}$ equal to the identity operator. The overall  dimension  ${\cal S}_z$ is thus $d N!\times d N!$  

\begin{equation}\label{switchN}
	{\mathcal S}(\N{1},\N{2},\ldots,\N{N}) \left( \rho \otimes \rho_c \right) = \sum_{z=0}^N {\cal S}_z,
\end{equation}
and (see Appendices \ref{AppA} and \ref{Rules})
\begin{equation} \label{Sz}
	{\cal S}_z = \sum_{k, k'=1}^{N!} \sqrt{P_k P_{k'}} \sum_{A_z\in {\bf A}_z^N}
	f_{A_z}\cdot
	Q^{k,k'}_{A_z} \otimes \left| k \right> \left< k' \right|
\end{equation}
with $$f_{A_z}=d^{2(z-N)}\prod_{j=1}^{N}( 1-q_j)\prod_{a\in A_z} \frac{q_{a}}{1-q_{a}}  
$$
\noindent where ${\bf A}_z^N$  is the collection of all possible subsets ${A}_z$ of $z$ subscripts in $\{1,2,\ldots,N\}$ corresponding to the $z$ indices  equal to zero (i.e. $i_a=0$ $\forall a \in A_z$). The following subsections detail examples with $N=2$ and $N=3$. The coefficients $ Q^{k,k'}_{A_z}$ are given by

\begin{equation}\label{Qkk}
	Q^{k,k'}_{A_z} = \sum_{\{i_b|b\in B_z\}}
	\pi_{k}\left( U_{i_{1}}\cdots U_{i_{N}}\right)\rho \left[ \pi_{k'}\left(U_{i_{1}}\cdots U_{i_{N}} \right) \right] ^\dagger.
\end{equation}

\noindent The $U_{i_j}^j$ of equation (\ref{depch2}) have been simplified in $U_{i_j}$. We can see from equation~(\ref{Qkk}) that the elements of the matrix ${\mathcal S}(\N{1},\N{2},\ldots,\N{N}) \left( \rho \otimes \rho_c \right)$ will always be linear combination of $\rho$ and $\mathbb{I}_1$, whatever $N$ channels.  Note also the operators $U_{i_j}$ for ${j\in A_z}$ are identity operators $\mathbb{I}_t$ by construction. Thus, arguments $U_{i_{1}}\cdots U_{i_{N}}$ under $\pi_k$ in (\ref{Qkk}) involves $N$ elements, $z$ of them in $A_z$ and $N-z$ in $B_z$. The matrix $\mathcal{S}$ and the pivotal equations~(\ref{switchN}-\ref{Qkk}) contain all information about the correlations between precise  causal orders coherently controlled by $\rho_c$ and the output of the quantum switch. $\mathcal{S}$ is a function of several parameters: the involved causal orders $\pi_k$ via the probabilities $P_k$, the depolarization strengths $q_i$'s of each individual channel $\N{i}$, the dimension  $d$ of the target system undergoing the operations of those channels and the number of channels $N$. Notably the sum over $k$ and $k'$ in equation (\ref{Sz}) can be restricted to a subset of definite causal orders via the probabilities $P_k$, i.e. a subset of superposition of $m$ causal orders among the $N!$ existing ones for advanced quantum control. This handle had remained unexplored up to now. It was not accessible to former explorations limited to two channels. In the current work we consider only superpositions of all causal orders. The control of causal orders will be presented elsewhere.

In the following subsections we will give the explicit expressions of the quantum switch matrices for the quantum $N$-switch channel for $N=2$ and $N=3$. We access these matrices of the quantum $N$-switch channel via the systematic ordering of the terms in equations (\ref{Krausg}) as settled in equations~(\ref{switchN}-\ref{Qkk}). 

The explicit calculation of the quantum $N$-switch channel gives important insights on the transmission of information coherently controlled by $\rho_c$ in a fascinating multi-parameter space. We briefly review below some of the intriguing behaviors associated to the parameters exploration in the $N=2$ and the $N=3$ cases. We show indeed in those cases how the nature and number of the causal orders in the control state superposition, the dimension of the target system, the level of noise all play a role. We underline that the $N=3$ case is still untouched experimentally.

To derive equation (\ref{Sz}) for particular cases of $N$, we first introduce the definitions of $W_{\mathbf{i}}$ and $\rho_c$ into equation (\ref{Krausg}). Introducing the definitions of  the Kraus operators in terms of $U_{i_j}^j$ operators and  applying the same reordering on the sums as in equation (\ref{sumABz})
leads to equation (\ref{Sz}).
In the following subsection, specific developments for $N=2$ and $N=3$ to evaluate the $Q^{k,k'}_{A_z}$ are given simplifying $Q^{k,k'}_{A_z}$ in (\ref{Qkk}) by following the relations presented in Appendix \ref{Rules} (equations (\ref{unitaryproperty1})-(\ref{unitaryproperty3})).

\section{The quantum switch matrices for $N=2$ and $N=3$}
\label{results}
To show the usefulness of equations (\ref{Sz}), we derive general expressions to investigate the transmission of information through two and three channels in an indefinite causal order. Our method can be easily applied to any number of depolarizing channels provided that $\{ U_{i} \}_{i=1}^{d^2}$ are unitary operators setting an orthonormal basis for the space of $d\times d$ matrices. 

\subsection{Evaluation of $\mathcal{S}$ for $N=2$}  \label{detailsN2} 

To explicitly evaluate equation (\ref{switchN}) with two channels, we identify the two permutations in $S_2$ : $\pi_1=\begin{psmallmatrix}1 & 2\\1 & 2\end{psmallmatrix}$ and $\pi_2=\begin{psmallmatrix}1 & 2\\2 & 1\end{psmallmatrix}$. Equation~(\ref{switchN}) for the quantum 2-switch channel matrix acting on the input state $\rho\otimes\rho_c$ writes 

\begin{equation}
	{\mathcal S}(\N{1},\N{2})(\rho\otimes\rho_c) =  {\cal S}_0 + {\cal S}_1 + {\cal S}_2 .  
\end{equation}

The collection of all subsets of subscripts in $\{1,2\}$ 
are ${\bf A}_0^2=\{\emptyset\}, {\bf A}_1^2=\{ \{ 1\}, \{ 2\} \}$ and ${\bf A}_2^2=\{ \{ 1,  2\} \}$. Then, the corresponding complementary collections are 

\noindent $ {\bf B}_0^2=\{ \{ 1,  2\} \}, {\bf B}_1^2=\{ \{ 2\}, \{ 1\} \}$ and ${\bf B}_2^2=\{\emptyset\}$. \\

\noindent \textit{Coefficients for $\mathcal{S}_0$}.  In this case, we use ${\bf A}_0^2=\{\emptyset\}$ to calculate the coefficients $Q^{k,k'}_{\emptyset}$, $k,k'\in \{1,2\}$. The  $Q^{k,k'}_{\emptyset}$ then reads 
\begin{align}
	\begin{array}{ll}
		Q^{1,1}_{\emptyset}= \sum_{i_1,i_2}\pi_1(U_{i_1} U_{i_2})\rho\pi_1(U_{i_1} U_{i_2} )^\dagger\\ 
		\hspace*{0.75cm} =\sum_{i_1,i_2}(U_{i_1} U_{i_2})\rho ( U_{i_2}^\dagger U_{i_1}^\dagger)\\
		\hspace*{0.75cm} =d\sum_{i_1,i_2} U_{i_1}  U_{i_1}^\dagger = d^3 \mathbb{I}.\\
		Q^{1,2}_{\emptyset}= \sum_{i_1,i_2}\pi_1(U_{i_1} U_{i_2})\rho\pi_2(U_{i_1} U_{i_2} )^\dagger\\
		\hspace*{0.75cm}=\sum_{i_1,i_2}(U_{i_1} U_{i_2})\rho (U_{i_1}^\dagger U_{i_2}^\dagger)\\
		\hspace*{0.75cm} =d\sum_{i_1,} U_{i_1} {\rm tr} (\rho U_{i_1}^\dagger )  = d^2 \rho.
	\end{array}
\end{align}

\noindent where we have used equations~(\ref{unitaryproperty1}) and (\ref{unitaryproperty3}) for $Q^{1,1}_{\emptyset}$, equation~(\ref{unitaryproperty1}) with $X= U_{i_2}\rho$  and  equation (\ref{unitaryproperty2})  for  $Q^{1,2}_{\emptyset}$.
\noindent Likewise, we have
$Q^{\alpha,\alpha'}_{\emptyset}=d^3\mathbb{I}, \quad \text{for} \quad (\alpha,\alpha')\in {\mathfrak A}\equiv\{(1,1), (2,2)\}$ and $ Q^{\beta,\beta'}_{\emptyset}=d^2 \rho, \quad \text{for} \quad  (\beta,\beta')\in {\mathfrak B}\equiv \{(1,2), (2,1) \}$. Then, we may write
\begin{equation}\label{2S0}
	{\cal S}_0=\displaystyle \sum_{(\alpha,\alpha')\in {\mathfrak A}}
	\frac{r_0  \mathbb{I}}{d}\sqrt{P_{\alpha}P_{\alpha'}}
	\otimes \vert \alpha \rangle\langle \alpha'\vert+ \sum_{(\beta,\beta')\in {\mathfrak B}}
	\frac{r_0 \rho}{d^2} \sqrt{P_{\beta}P_{\beta'}}
	\otimes \vert \beta\rangle\langle \beta' \vert, 
\end{equation}
\noindent where $r_0 = p_1p_2$ with $p_i=1-q_i$. \\

\noindent \textit{Coefficients for $\mathcal{S}_1$}. In this case ${\bf A}_1^2=\{\{1\},\{2\}\}$ and ${\bf B}_1^2=\{\{2\},\{1\}\}$.
Let us first consider the coefficient $Q^{\gamma,\gamma'}_{\{1\}}=\sum_{i_2}  \pi_\gamma ( \mathbb{I} \cdot U_{i_2}) \rho \pi_{\gamma'} (\mathbb{I} \cdot U_{i_2})^\dagger = d \mathbb{I}$, using the general relations (\ref{unitaryproperty1})-(\ref{unitaryproperty3}), 
for $(\gamma,\gamma')\in {\mathfrak G} \equiv \{(1,1), (1,2), (2,1), (2,2)\}$. 
Since indices are dumb it can be shown that $Q^{\gamma,\gamma'}_{\{2\}}=Q^{\gamma,\gamma'}_{\{1\}}$ for all $(\gamma, \gamma')$. 
Then the term ${\cal S}_1$ can be written as

\begin{equation}\label{2S1}
	{\cal S}_1=\sum_{k,k'} \displaystyle\frac{r_1}{d}\sqrt{P_kP_{k'}}  \mathbb{I} \otimes \vert k\rangle\langle k'\vert=\frac{r_1}{d} \mathbb{I}\otimes \rho_c, 
\end{equation}
\noindent where $r_1=q_1p_2+q_2p_1. $ 

\noindent {\it Coefficients for $\mathcal{S}_2$. }Finally, let us consider the term ${\cal S}_2$. In this case ${\bf A}_2^2=\{\{1,2\}\}$ and hence ${ {\bf B}_2^2}=\{\emptyset\}$. Note that $Q^{k,k'}_{\{1,2\}}=\rho$ for all $k$ and  $k'$. Thus, the term with $z=2$ reads
\begin{equation}\label{2S2}
	{\cal S}_2 =   \sum_{k,k'} r_2  \rho \sqrt{P_k P_{k'}} \otimes \vert k\rangle\langle k'\vert= {r_2} \rho\otimes \rho_c,
\end{equation}
with $r_2=q_1q_2$.
By expanding the matrices $\mathcal{S}_0$, $\mathcal{S}_1$ and $\mathcal{S}_2$ in the control qubit basis, $\{ \left| 1\right>,  \left|2\right>\}$, we are able to write  
\begin{equation} \label{Matrices2}
	\begin{array}{lll}
		{\mathcal S}_0=\left(\begin{array}{cc} \frac{r_0}{d} \mathbb{I} P_1 &\frac{r_0 \rho}{d^2} \sqrt{P_1 P_2} \\ 
			\frac{r_0 \rho}{d^2} \sqrt{P_2 P_1} &\frac{r_0}{d} \mathbb{I} P_2  \end{array}\right), \\ [1em]
		{\mathcal S}_1=\left(\begin{array}{cc} \frac{r_1 }{d} \mathbb{I}  P_1 &\frac{r_1 }{d} \mathbb{I} \sqrt{P_1P_2}\\ 
			\frac{r_1 }{d} \mathbb{I} \sqrt{P_2P_1} &\frac{r_1 }{d} \mathbb{I} P_2 \end{array}\right),\\ [1em]
		{\mathcal S}_2=\left(\begin{array}{cc} r_2  \rho P_1 &r_2  \rho \sqrt{P_1 P_{2}}\\ 
			r_2  \rho \sqrt{P_2 P_{1}} &r_2  \rho P_2 \end{array}\right).
	\end{array}
\end{equation} 
\noindent where $\mathbb{I}=\mathbb{I}_t$. 
Summing those matrices according to equation~(\ref{switchN}), we find that the quantum 2-switch channel matrix ${\mathcal S}(\N{1},\N{2})$ has diagonal elements $ a_k=P_k[(r_0+r_1) \mathbb{I}/d+r_2\rho], $ for $ k=1,2 $ and  off-diagonal elements $b=\sqrt{P_1P_2}[(r_0+d^2r_2)\rho/d^2+\frac{r_1}{d} \mathbb{I}] $, with $r_0=p_1p_2$, $r_1=q_1p_2+q_2p_1 $ and $r_2=q_1q_2$. Thus,
\begin{equation}\label{Matrix2}
	\begin{array}{ll}
		{\mathcal S}(\N{1},\N{2})(\rho \otimes \rho_c)=\left(\begin{array}{cc}
			a_1 &b \\ 
			b &a_2\end{array}\right),
	\end{array}
\end{equation}

\noindent note that the diagonal and off-diagonal elements $a_1$, $a_2$ and $b$ are matrices and are linear combinations of matrices $\rho$ and $\mathbb{I}_t$. This property is non-unique for case $N=2$, instead is general for $N$ channels, an advisable aspect from equation (\ref{Qkk}) and equations (\ref{unitaryproperty1})-(\ref{unitaryproperty2}). Indeed, (\ref{Matrix2}) gives as  particular outputs the predicted Holevo capacity of Figure 3 in \cite{goswami2018communicating} and expressions of Holevo information in \cite{ebler2018enhanced}. 

We end up this subsection stressing that Figure.~\ref{Figura1} sketches different ways to connect channels $\N{1}$ and $\N{2}$ in either (a) and (b) a definite causal order and (c) for an indefinite causal order combining the $2!$ possible orders.

\subsection{Evaluation of $\mathcal{S}$ for $N=3$}\label{detailsN3} 
In this section, we explicitly evaluate expression~(\ref{switchN}) considering three channels. Let us label the 6 elements of $S_3$ according to the following set of permutations
$\pi_1=\begin{psmallmatrix} 1 &2 & 3\\ 1&2&3\end{psmallmatrix}$, $\pi_2=\begin{psmallmatrix}1 &2 & 3\\ 1&3&2\end{psmallmatrix}$, $\pi_3= \begin{psmallmatrix}1&2&3\\ 2 &1 & 3\end{psmallmatrix}$, $\pi_4=\begin{psmallmatrix} 1 &2 & 3\\ 2&3&1\end{psmallmatrix}$, $\pi_5= \begin{psmallmatrix}1 &2 & 3\\ 3&1&2\end{psmallmatrix}$ and $\pi_6=\begin{psmallmatrix} 1 &2 & 3\\ 3&2&1\end{psmallmatrix}$. Equation (\ref{switchN}) for the quantum 3-switch channel matrix acting on input state $\rho\otimes\rho_c$ reads 
\begin{equation}
	{\mathcal S}(\N{1},\N{2},\N{3}) \left( \rho \otimes \rho_c \right) =  {\cal S}_0 + {\cal S}_1 + {\cal S}_2 +{\cal S}_3.
\end{equation}

\noindent \textit{Coefficients for $\mathcal{S}_0$}. In this case note that ${\bf A}_0^3=\{\emptyset\}$, hence ${{\bf B}_0^3}=\{\{1,2,3\}\}$. Besides, the sum in $Q^{1,k'}_{\emptyset}$ is over the indices $\{i_1,i_2, i_3\}$. These can be computed explicitly 
\begin{equation}
	Q^{1,1}_{\emptyset}= \sum_{i_1,i_2,i_3}\pi_1(U_{i_1} U_{i_2} U_{i_3})\rho\pi_1(U_{i_1} U_{i_2} U_{i_3})^\dagger= d^5 \mathbb{I}.
\end{equation}
Likewise,
\begin{equation}
	Q^{1,4}_{\emptyset}= \sum_{i_1,i_2,i_3} \pi_1(U_{i_1} U_{i_2}U_{i_3})\rho \pi_4(U_{i_1} U_{i_2}U_{i_3})^\dagger=d^4 \rho.
\end{equation} 
The remaining coefficients for ${\cal S}_0$ are 
\begin{equation}\label{Q140}
	\begin{array}{ll}
		Q^{1,2}_{\emptyset}= \sum_{i_1,i_2,i_3}\pi_1(U_{i_1} U_{i_2}U_{i_3})\rho\pi_2(U_{i_1} U_{i_2} U_{i_3})^\dagger\\
		\hspace*{.75cm}=\sum_{i_1,i_2,i_3}(U_{i_1} U_{i_2}U_{i_3})\rho (U_{i_2}^\dagger U_{i_3}^\dagger U_{i_1}^\dagger)\\
		\hspace*{0.75cm}=d\sum_{i_1,i_3} U_{i_1} \tr ( U_{i_3} \rho) U_{i_3}^\dagger U_{i_1}^\dagger=d^2\sum_{i_1}  U_{i_1} \rho U_{i_1}^\dagger\\
		\hspace*{0.75cm} = d^3 \mathbb{I}, \\[0em]
		Q^{1,3}_{\emptyset}= \sum_{i_1,i_2,i_3}\pi_1(U_{i_1} U_{i_2}U_{i_3})\rho\pi_3(U_{i_1} U_{i_2}  U_{i_3} )^\dagger\\
		\hspace*{0.75cm}=\sum_{i_1,i_2,i_3}(U_{i_1} U_{i_2}U_{i_3})\rho (U_{i_3}^\dagger U_{i_1}^\dagger U_{i_2}^\dagger)\\
		\hspace*{0.7cm} =d\sum_{i_1,i_2} U_{i_1} U_{i_2}\mathbb{I}  U_{i_1}^\dagger U_{i_2}^\dagger=d^2\sum_{i_1}  \tr (U_{i_2} \mathbb{I})  U_{i_2}^\dagger\\
		\hspace*{0.75cm} = d^3 \mathbb{I},\\[0em]
		Q^{1,5}_{\emptyset}= \sum_{i_1,i_2,i_3}\pi_1(U_{i_1} U_{i_2}U_{i_3})\rho\pi_5(U_{i_1} U_{i_2} U_{i_3})^\dagger\\
		\hspace*{0.75cm}=\sum_{i_1,i_2,i_3}(U_{i_1} U_{i_2}U_{i_3})\rho (U_{i_2}^\dagger U_{i_1}^\dagger U_{i_3}^\dagger)\\
		\hspace*{0.75cm} =d\sum_{i_1,i_3} U_{i_1} \tr ( U_{i_3} \rho) U_{i_1}^\dagger U_{i_3}^\dagger=d^3\sum_{i_3} \tr (U_{i_3} \rho) U_{i_3}^\dagger\\
		\hspace*{0.75cm} = d^4 \rho,\\[0em]
		Q^{1,6}_{\emptyset}= \sum_{i_1,i_2,i_3}\pi_1(U_{i_1} U_{i_2}U_{i_3})\rho\pi_6(U_{i_1} U_{i_2} U_{i_3})^\dagger\\
		\hspace*{0.75cm}=\sum_{i_1,i_2,i_3}(U_{i_1} U_{i_2}U_{i_3})\rho (U_{i_1}^\dagger U_{i_2}^\dagger U_{i_3}^\dagger)\\
		\hspace*{0.75cm} = d\sum_{i_1,i_3} U_{i_1} \tr (U_{i_3} \rho U_{i_1}^\dagger) U_{i_3}^\dagger=d^2 \sum_{i_1} U_{i_1}\rho U_{i_1}^\dagger\\
		\hspace*{0.75cm}=d^3 \mathbb{I}.
	\end{array}
\end{equation}
\noindent The coefficients $Q^{k,k'}_{\emptyset}$ with $k\ge 2$ can be computed using these expressions from equations (\ref{Q140}). For instance, consider the following 
\begin{equation}
	Q^{2,6}_{\emptyset}= \sum_{i_1,i_2,i_3}\pi_2(U_{i_1} U_{i_2}U_{i_3})\rho\pi_6(U_{i_1} U_{i_2} U_{i_3})^\dagger=\sum_{i_1,i_2,i_3}(U_{i_1} U_{i_3}U_{i_2})\rho (U_{i_1}^\dagger U_{i_2}^\dagger U_{i_3}^\dagger),
\end{equation}
which is equivalent to expression $Q^{1,4}_{\emptyset}$ because the indices $i$'s are dumb.  Thus one can calculate explicitly the remaining coefficients. Results are thus summarized in the following list
\begin{equation}
	\begin{array}{ll}
		Q^{i,i'}_{\emptyset}=d^3\mathbb{I}, \forall \hspace*{0.1cm}  (i,i')\in {\mathfrak I}\equiv \{(1,6), (2,4), (3,5), (4,2),\\
		\hspace*{3.5cm}  (1,2),(2,1), (3,4), (4,3), (5,6),  \\
		\hspace*{3.5cm}  (6,5), (5,3),(6,1), (1,3), (2,5), \\
		\hspace*{3.5 cm}   (3,1), (4,6), (5,2),(6,4)\},\\
		Q^{j,j'}_{\emptyset}=d^4\rho, \forall \hspace*{0.1cm}  (j,j')\in {\mathfrak J}\equiv\{(1,4), (2,6), (3,2), (4,5),   \\
		\hspace*{3.5cm} (5,1),(6,3), (1,5), (2,3), (3,6), \\
		\hspace*{3.5cm} (4,1), (5,4),(6,2)\},\\
		Q^{k,k'}_{\emptyset}=d^5\mathbb{I}, \forall \hspace*{0.1cm}  (k,k')\in {\mathfrak K}\equiv\{(1,1), (2,2), (3,3),\\
		\hspace*{3.5cm} (4,4), (5,5),(6,6)\}.\\
	\end{array}
\end{equation}
\noindent After calculating all these coefficients, we obtain
\begin{equation}\label{S0}
	\begin{array}{ll}
		{\cal S}_0=\displaystyle \sum_{(i,i')\in {\mathfrak I}} \frac{s_0 }{d^3} \mathbb{I}\sqrt{P_iP_{i'}}\otimes \vert i\rangle\langle i'\vert 
		%
		+ \sum_{(j,j')\in {\mathfrak J}} \frac{s_0 \rho}{d^2}\sqrt{P_jP_{j'}}\otimes \vert j \rangle \langle j' \vert\\[1.5em]
		\hspace*{6cm}\displaystyle 
		+\sum_{(k,k')\in {\mathfrak K} } \frac{s_0 }{d} \mathbb{I} \sqrt{P_kP_{k'}}\otimes \vert k\rangle\langle k'\vert, 
	\end{array}
\end{equation}

\noindent where $s_0=p_1p_2p_3$. \\

\noindent \textit{Coefficients for $\mathcal{S}_1$.} In this case ${\bf A}_1^3=\{\{1\},\{2\},\{3\}\}$ and ${{\bf B}_1^3}=\{\{2,3\},\{1,3\},\{1,2\}\}$.
Let us first consider the coefficient $Q^{k,k'}_{\{1\}}$, so that sum must be accomplished over the indices $\{i_2,i_3\}$, hence $Q^{k,k'}_{\{1\}}=\sum_{i_2,i_3}\pi_k ( \mathbb{I} \cdot U_{i_2} \cdot U_{i_3}) \rho \pi_{k'} (\mathbb{I} \cdot U_{i_2} \cdot U_{i_3})^\dagger$.
\noindent  Using the  relations~(\ref{unitaryproperty1})-(\ref{unitaryproperty3}) we obtain

\begin{equation}
	\begin{array}{ll}
		Q^{\ell,\ell'}_{\{1\}}=d^2 \rho,\forall \hspace*{0.1cm}  (\ell,\ell')\in {\mathfrak L}_1 \equiv \{(2,3), (3,2), (2,4), (4,2),  \\
		\hspace*{4cm}  (3,5), (5,3), (3,6),(6,3),(4,5),\\
		\hspace*{4cm} (5,4),(4,6),(6,4), (5,1), (1,5),  \\
		\hspace*{4cm} (1,2), (2,1),(1,6), (6,1)\},\\
		Q^{m,m'}_{\{1\}}=d^3\mathbb{I},\forall \hspace*{0.1cm} (m,m')\in {\mathfrak M}_1 \equiv \{(1,1), (2,2), (3,3), (4,4), \\
		\hspace*{4cm} (5,5), (6,6), (1,3),(1,4),(4,1)\\
		\hspace*{4cm} (3,1), (2,5),(5,2), (2,6), (6,2), \\
		\hspace*{4cm} (3,4), (4,3),(5,6),(6,5) \},\\
	\end{array}
\end{equation}

\begin{equation}
	\begin{array}{ll}
		Q^{\ell,\ell'}_{\{2\}}=d^2 \rho,\forall \hspace*{0.1cm}  (\ell,\ell')\in {\mathfrak L}_2 \equiv \{(1,4), (1,5), (1,6), (2,4), \\
		\hspace*{4cm}  (2,5), (2,6),(3,4),(3,5),(3,6)\\
		\hspace*{4cm} (4,1),(4,2),(4,3), (5,1), (5,2),  \\
		\hspace*{4cm} (5,3), (6,1),(6,2), (6,3)\},\\
		Q^{m,m'}_{\{2\}}=d^3\mathbb{I},\forall \hspace*{0.1cm} (m,m')\in {\mathfrak M}_2 \equiv \{(1,1), (1,2), (1,3), (2,1), \\
		\hspace*{4cm} (2,2), (2,3), (3,1),(3,2),(3,3)\\
		\hspace*{4cm} (4,4), (4,5),(4,6), (5,4), (5,5), \\
		\hspace*{4cm} (5,6), (6,4),(6,5),(6,6) \},\\
	\end{array}
\end{equation}


\begin{equation}
	\begin{array}{ll}
		Q^{\ell,\ell'}_{\{3\}}=d^2 \rho,\forall \hspace*{0.1cm}  (\ell,\ell')\in {\mathfrak L}_3 \equiv \{(1,3), (1,4), (1,6), (2,3), \\
		\hspace*{4cm}  (2,4), (2,6),(3,1),(3,2),(3,5)\\
		\hspace*{4cm} (4,1),(4,2),(4,5), (5,3), (5,4),  \\
		\hspace*{4cm} (5,6), (6,1),(6,2), (6,5)\},\\
		Q^{m,m'}_{\{3\}}=d^3\mathbb{I},\forall \hspace*{0.1cm} (m,m')\in {\mathfrak M}_3 \equiv \{(1,1), (1,2), (1,5), (2,1), \\
		\hspace*{4cm} (2,2), (2,5), (3,3),(3,4),(3,6)\\
		\hspace*{4cm} (4,3), (4,4),(4,6), (5,1), (5,2), \\
		\hspace*{4cm} (5,5), (6,3),(6,4),(6,6) \}.\\
	\end{array}
\end{equation}

Hence, the matrix ${\cal S}_1$ can be computed

\begin{equation}\label{S1}
	\begin{array}{ll}
		{\cal S}_1=\displaystyle\frac1{d^2} \sum_{s=1}^3 \left( t_s d \sum_{(\ell,\ell')\in {\mathfrak M}_s} \sqrt{P_\ell P_{\ell'}}\,\mathbb{I}\otimes \vert \ell \rangle \langle \ell'\vert\right. 
		%
		\left. +t_s \displaystyle \sum_{(m,m')\in {\mathfrak L}_s} \sqrt{P_m P_{m'}} \, \rho \otimes \vert m \rangle \langle m'\vert\right),
	\end{array}
\end{equation}
where $t_1=p_2p_3q_1$, $t_2=p_1p_3q_2$ and $t_3=p_1p_2q_3$. \\

\noindent \textit{Coefficients for $\mathcal{S}_2$.} In this case ${\bf A}_2^3=\{\{1,2\},\{1,3\},\{2,3\}\}$ and hence 
${ {\bf B}_2^3}=\{\{3\},\{2\},\{1\}\}$. Let us consider 
\begin{equation*}
	Q^{k,k'}_{\{2,3\}}=\sum_{i_1}\pi_k (U_{i_1} \cdot\mathbb{I} \cdot \mathbb{I}) \rho \pi_{k'} (U_{i_1} \cdot \mathbb{I}\cdot \mathbb{I})^\dagger=d \mathbb{I},
\end{equation*}
where the operators $\mathbb{I}$ have been written for the sake of clarity as the permutations $\pi_k$ act on sets of three elements. In a similar way $Q^{k,k'}_{\{1,3\}}=Q^{k,k'}_{\{1,2\}}=d \mathbb{I}.$
Thus, we obtain
\begin{equation}\label{S2}
	\begin{array}{ll}
		{\cal S}_2= \frac{p_1p_2p_3}{d^2} \sum_{k,k'} \sqrt{P_kP_{k'}} \left(\frac{q_2q_3}{p_2p_3} Q^{k,k'}_{\{2,3\}} \right.\\[1em]
		\hspace*{1cm} + \left. \frac{q_1q_3}{p_1p_3} Q^{k,k'}_{\{1,3\}} + \frac{q_1q_2}{p_1p_2} Q^{k,k'}_{\{1,2\}} \right) \otimes \vert k  \rangle \langle k' \vert= \frac{s_2}{d} \mathbb{I} \otimes \rho_c
	\end{array}
\end{equation}
\noindent where  $s_2= q_1q_2p_3+q_1q_3p_2+q_2q_3p_1$.\\ 

\noindent \textit{Coefficients for $\mathcal{S}_3$.} Finally, note that $Q^{k,k'}_{\{1,2,3\}}=\rho$ for all $k$ and $k'$. Thus, the term with $z=3$ reads
\begin{equation}\label{S3}
	{\cal S}_3 =  s_3 \sum_{k,k'} \sqrt{P_kP_{k'}} \rho \otimes \vert k\rangle \langle k' \vert= {s_3} \rho\otimes \rho_c, 
\end{equation}
\noindent where $s_3=q_1q_2q_3$ and using the definition of the control qudit.

For three channels, Figure~\ref{Figura2} shows different ways to connect channels $\N{1}$, $\N{2}$  and $\N{3}$ in either (a)-(f) a definite causal order, or (g) in an indefinite causal order taking into account all 3! causal orders. The quantum 3-switch  matrix is again calculated with equation (\ref{switchN}) (see Appendix~\ref{AppC})

\begin{equation}\label{Matrix3i}
	\begin{array}{lll}
		{\mathcal S}(\N{1},\N{2},\N{3})(\rho \otimes \rho_c)=\left(\begin{array}{cccccc} 
			\mathcal{A}_1 &\mathcal{B} & \mathcal{C}&\mathcal{D}&\mathcal{E}&\mathcal{F}\\
			\mathcal{B} &\mathcal{A}_2 &\mathcal{G}&\mathcal{H}&\mathcal{I}&\mathcal{J}\\ 
			\mathcal{C} &\mathcal{G} & \mathcal{A}_3&\mathcal{K}&\mathcal{L}&\mathcal{M}\\ 
			\mathcal{D} &\mathcal{H} & \mathcal{K}&\mathcal{A}_4&\mathcal{N}&\mathcal{P}\\
			\mathcal{E} &\mathcal{I}& \mathcal{L}&\mathcal{N}&\mathcal{A}_5&\mathcal{Q}\\
			\mathcal{F}&\mathcal{J} & \mathcal{M}&\mathcal{P}&\mathcal{Q}&\mathcal{A}_6\end{array}\right),
	\end{array}
\end{equation}

\noindent where the diagonal and the off-diagonal elements whose expressions are given in Appendix~\ref{AppC} are also linear combinations of matrices $\rho$ and $\mathbb{I}_t$. From the definition of symmetric matrices \cite{horn1990matrix}, we can see that the quantum switch matrices~(\ref{Matrix2}) and (\ref{Matrix3i}) are block-symmetric matrices with respect to the main diagonal. This could be seen as general from the fact $Q^{k,k'}_{A_z}=Q^{k',k}_{A_z}$ due to equations (\ref{Sz}) and (\ref{Qkk}), because indices in the sums are dumb. Thus, as the number of channels increases, the number of different $d\times d$ matrices involved in the quantum $N$-switch matrix $\mathcal{S}$ scales as $N!(N!+1)/2$. Notice that those matrices also characterize information transmission of any definite causal ordering $\pi_k$ of channels $\mathcal{N}_{\pi_k}$ when setting $P_k=1$ and $P_s=0$ for all $s\neq k$. 

Matrices in equation (\ref{Matrix2}) or (\ref{Matrix3i}) are written in the basis of the control system $\rho_c$ which maps and weights the chosen causal orders. To know the best rate to communicate classical information with two and three channels, in the following Section we diagonalize matrices~(\ref{Matrix2}) and (\ref{Matrix3i}) to compute the Holevo information limit $\chi$, which quantifies how much classical information can be transmitted through a channel  in a single use. $\chi$  gives a lower bound on the classical capacity \cite{holevo1998capacity, abbott2018communication,schumacher1997sending}. 

\section{Holevo information limit for two and three channels}\label{Holevol23}

We compute the Holevo information limit (Holevo information for shortness in the following) $\chi(\mathcal{S})$ for $N=2$ and $N=3$ channels through a generalization of the mutual information (see for example \cite{wilde2013quantum}) and supplementary information of \cite{ebler2018enhanced}. The Holevo information $\chi(\mathcal{S})$ is found by maximizing mutual information, and it can be shown that maximization over the $\rho$ pure states is sufficient \cite{wilde2013quantum}. The Holevo information is then given by
\begin{equation}\label{Gholevo}
	\begin{array}{ll}
		\QNS \big({\mathcal S}\big) = \log d + H({\tilde \rho}_c^{(N)}) - H^\text{min}({\mathcal S})
	\end{array}
\end{equation}

\noindent where $d$ is the dimension of the target system $\rho$, $H({\tilde \rho}_c^{(N)})$ is the von-Neumann entropy of the output control system $\tilde \rho_c^{(N)}$ for $N$ channels and  $H^\text{min}(\mathcal{S})$ is the minimum of the entropy at the output  of the channel $\mathcal{S}$. The minimization of $H^\text{min}(\mathcal{S})\equiv \underset{\rho}{\text{min}}  \hspace{1ex} H^\text{min}(\mathcal{S}(\rho))$ is over all input states $\rho$ going on the channel $\mathcal{S}$ \cite{wilde2013quantum}.  
To evaluate equation~(\ref{Gholevo}):  
\begin{enumerate}
	\item The diagonalization and minimization of $H^\text{min}({\mathcal S})$ is performed on all possible states given by $\rho$. It is done analytically for $N=2$ channels and arbitrary $q_i$. For $N=3$ channels we compute the eigenvalues of the full quantum 3-switch matrix  $\mathcal{S}(\N{1},\N{2},\N{3} )\left( \rho \otimes \rho_c \right)$  numerically. 
	\item ${\tilde \rho}_c^{(N)}$ was analytically calculated following \cite{ebler2018enhanced}.
	\item We deduce $H({\tilde \rho}_c^{(N)})$ from the analytical expressions of ${\tilde \rho}_c^{(N)}$. 
\end{enumerate}

\subsection{Holevo information limit for $N=2$ channels}

\subsubsection{Calculation of $H^\text{min}$}\label{CalculationHmin} 

\noindent We calculate the minimum output entropy $H^\text{min}(\mathcal{S})$ of the channel $\mathcal{S}\equiv{\mathcal S}(\N{1},\ldots,\N{N})$
\begin{equation*}
	H^\text{min}(\mathcal{S})\equiv \underset{\rho}{\text{min}}  \hspace{1ex} H^\text{min}(\mathcal{S}(\rho))=\underset{\rho}{\text{min}}\sum_i-\lambda_{{\cal S}(\rho),i}\text{log}[\lambda_{{\cal S}(\rho),i}], 
\end{equation*}
where the minimization is a priori over all input states $\rho$ and $\{\lambda_{{\cal S}(\rho),i}\}_{i=1}^d$  are the eigenvalues of ${\cal S}(\rho)$. In fact it is sufficient to minimize over the states \cite{wilde2013quantum} and the eigenvalues $\{\lambda_{\rho,i}\}_{i=1}^d$ sum up to $1$. As $H^\text{min}(\mathcal{S}(\rho))$ is concave, the minimization is done as in Ref. \cite{ebler2018enhanced} : the eigenvalues $\{\lambda_{{\cal S}(\rho),i}\}_{i=1}^d$ are taken at the border of the interval $[0,1]^{\times d}$ and as they sum up to one, the minimization is simplified to the cases where all $\lambda$ but one are set to zero and the last one is equal to 1. 

In this situation, ${\mathcal S}(\N{1},\N{2}) \left( \rho \otimes \rho_c \right)$ has only four non-zero matrix elements, (see equation~(\ref{Matrix2})), which can be rewritten as $2\times2$ matrices 
\begin{equation}
	\left(\begin{matrix}
		a_0 p & b \cr 
		b & a_0 q
	\end{matrix}\right)
\end{equation}
\noindent where   $a_0$ and $b$  are  $d\times d$ matrices and linear combinations of $\rho$ and $\mathbb{I}_t$: 
\begin{equation}
	a_0=(r_0+r_1) \mathbb{I}/d+r_2\rho, \quad b=\sqrt{P_1P_2}[(r_0+d^2r_2)\rho/2+r_1\mathbb{I}/d],
\end{equation}
with $p\equiv P_1$, $q\equiv P_2$ are the control probabilities with $p+q=1$. 

Using the commutativity of $\rho$ and $\mathbb{I}_t$ (so they have the same eigenvectors), we then retrieve analytically $a_\pm$, the matrix-eigenvalues of ${\mathcal S}(\N{1},\N{2},\ldots,\N{N}) \left( \rho \otimes \rho_c \right)$  
\begin{equation}\label{solutionsas}
	\quad a_\pm = \frac{a_0}{2} \pm \sqrt{b^2 + a_0^2 (p - \frac{1}{2})^2}. 
\end{equation}
\noindent The existence of this last expression is warranted by the positivity of the discriminant \cite{bhatia2009positive}, considering the positivity of $\rho$ and the structure of $a_0$ and $b$, which are linear combinations of $\mathbb{I}$ and $\rho$.

The commutativity properties of $\rho$ and $\mathbb{I}_t$ are inherited to $a_\pm$. Then, the eigenvalues of ${\mathcal S}(\N{1},\N{2}) \left( \rho \otimes \rho_c \right)$ for two causal orders are the eigenvalues of $a_\pm$.
Thus, to diagonalize $a_\pm$ we just replace $\rho$ by its eigenvalues, labeled as $\lambda_{\rho,i}$, in equation (\ref{solutionsas}), which generalizes the procedure obtained in \cite{ebler2018enhanced}. Our procedure gives access to the transmission of information in a more general situation, where the depolarization strengths $q_i$ can be different for each channel and it can take any value between 0 and 1. 
Equation (\ref{solutionsas}) gives the eigenvalues of the matrix ${\mathcal S}(\N{1},\N{2})(\rho \otimes \rho_c)$ ($s=\pm1$)
\begin{eqnarray}
	\lambda_{s,i} &=& \frac{\alpha_0}{2}+ s \sqrt{p q \beta^2 + \alpha_0^2 (p -\frac{1}{2})^2} \\
	&\text {with:}& \hspace{1ex}\alpha_0 \equiv \frac{1 - q_1 q_2}{d}  + q_1 q_2 \lambda_{\rho,i} \nonumber \\
	&& \beta \equiv \frac{p_1 q_2 + q_1 p_2}{d} + (\frac{p_1 p_2}{d^2}+q_1 q_2) \lambda_{\rho,i} \nonumber
\end{eqnarray}
\noindent The eigenvalues of $\lambda_{s,i}$ are well defined because of the positivity of discriminant \cite{bhatia2009positive}. Finally, using the concavity of the entropy, the minimum of the entropy $H^{\rm min}$ for a state is reached by setting just one $\lambda_{\rho,i}$ to one and all the others to zero, with this we obtain

\begin{equation}\label{Hmin}
	- H^\text{min}({\mathcal S}(\N{1},\N{2})) = \sum_{\substack{{s \in \{\pm 1\}}\\ {k \in \{ 0, 1 \}} }}  (d-1)^{1-k} \lambda_{s,k} \log \left( \lambda_{s,k} \right)  
\end{equation}

\begin{equation}
	\lambda_{s,k} =  \frac{{\alpha_{0,k}}}{2} + s \sqrt{p q \beta_k^2 + \alpha_{0,k}^2 (p -\frac{1}{2})^2}
\end{equation} 
\begin{equation}
	\alpha_{0,k} = \frac{1 - q_1 q_2}{d} + {k q_1 q_2}
	\label{alfaq}
\end{equation}
\begin{equation}
	\beta_{k} = \frac{p_1 q_2 + q_1 p_2}{d} + k \left( \frac{p_1 p_2}{d^2}+q_1 q_2 \right)
\end{equation}
\noindent It is easy to show $\beta_k \le {\alpha_{0,k}}$, then $\lambda_{\pm,i} \ge 0$ $(\lambda_{s,k} \ge 0)$ as expected. Also, $0 \le \lambda_{s,k} \le 1$ and then $- H^{\rm min}(\mathcal{S}({\mathcal N}_1,{\mathcal N}_2)) \le 0$. 

\begin{figure} [h!]
	\vspace*{13pt}
	\scalebox{.2}{\includegraphics{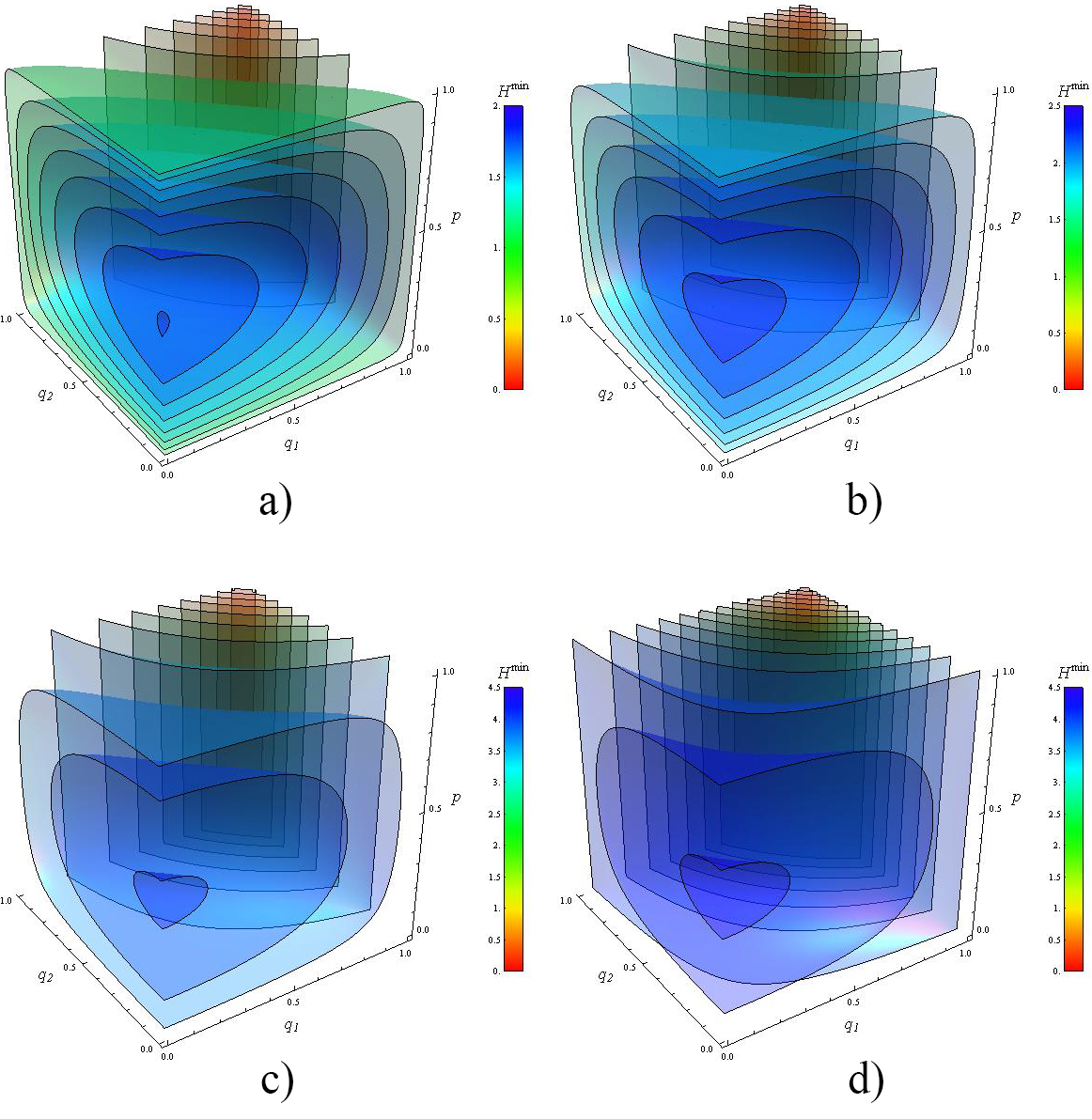}}
	\vspace*{13pt}
	\caption{\label{Entropy} {Entropy map for two noisy channels.} The 3D graphs represent contour surfaces of the Von-Neumann entropy  $H^{\rm min}(\mathcal{S}_2)$ when the depolarizing parameters $q_1, q_2,$ and the probabilities $P_1=P_2=p$ are varied from 0 to 1. We plot several cases when the dimension $d$ of the target $\rho$ is: a) $d=2$, b) $d=3$, c) $d=10$ and d) $d=100$. The value of $H^{\rm min}(\mathcal{S}_2)$ is depicted by the color in the bar besides. } 
\end{figure}

If one of $q_1=1$ , i.e. channel $1$ is free of depolarization, then $\alpha_{0,k}=\frac{p_2}{d}+k q_2=\beta_{k}$ and 
\begin{equation}\label{HminN21}
	- H^{\rm min}(\mathcal{S}({\mathcal N}_1,{\mathcal N}_2))=(d-1) \frac{p_2}{d} \log(\frac{p_2}{d}) + (\frac{p_2}{d}+q_2) \log(\frac{p_2}{d}+q_2) 
\end{equation}
depends only on the probability of depolarization for channel $2$. Thus, $- H^{\rm min}(\mathcal{S}({\mathcal N}_1,{\mathcal N}_2))$ reaches its maximum value of zero only if $q_1=q_2=1$. 
Alternatively, it is direct to show that the discriminant reaches its maximum value when $p=\frac{1}{2}$, which is the case studied by Ebler {\it et al.} \cite{ebler2018enhanced}. In addition, if $q_1=q_2=0$, i.e. both channels are fully depolarizing, then $\alpha_{0,k}=\frac{1}{d}, \beta_{k}=\frac{k}{d^2}$, so $- H^{\rm min}(\mathcal{S}({\mathcal N}_1,{\mathcal N}_2))$ reaches the minimum value
\begin{equation}\label{HminN2}
	\begin{array}{ll}
		- H^{\rm min}(\mathcal{S}({\mathcal N}_1,{\mathcal N}_2))= - \log(2d) + \frac{1}{2d^2} \log(\frac{d+1}{d-1})
		+ \frac{1}{2d} \log(1 - \frac{1}{d^2}). 
	\end{array}
\end{equation}
For sake of shortness the entropy $H^{\rm min}(\mathcal{S}({\mathcal N}_1,{\mathcal N}_2))$ will be denoted simply as $H^{\rm min}(\mathcal{S}_N)$. To illustrate the range of parameters of equation~\ref{HminN21}, we plot  the entropy $H^{\rm min}(\mathcal{S}_2)$ map for two noisy channels. Figure \ref{Entropy} shows the entropy $H^{\rm min}(\mathcal{S}_2)$. The plots are contour surfaces of $H^{\rm min}(\mathcal{S}_2)$ when $q_1,q_2$  vary from 0 to 1. Each plot contains thirty surfaces distributed in their complete range shown in the color-chart. We plot several cases of $H^{\rm min}(\mathcal{S}_2)$ when the  dimension of the target is $d=2, 3, 10$ and 100. 

\subsubsection{Derivation of ${\tilde \rho}_c^{(2)}$}\label{Rhocal}

To obtain the output state of the control system ${\tilde \rho}_c^{(N)}$ after $N$ channels, we calculate  
\begin{equation*}
	{\rm Tr}_{XIJ} \left[ (\mathcal{S}(\N{1},\ldots,{\mathcal N}_{N} )\left( \rho \otimes \rho_c \right)) \otimes \mathbb{I})(\omega_{XIJAC})\right] 
\end{equation*}
where $\omega_{XIJAC}$ is an extended input state with pure conditional state as described in \cite{ebler2018enhanced}. A direct calculation shows:
\begin{align}
	{\rm Tr}_{XIJ} & \left[ (\mathcal{S}(\N{1},\ldots,{\mathcal N}_{N} )\left( \rho \otimes \rho_c \right)) \otimes \mathbb{I})(\omega_{XIJAC}) \right] = \nonumber \\
	& = {\rm Tr}_{XIJ} \left[ \frac{1}{d^2} \sum_{x,i,j} p_x \left| x \left>\right< x \right| \left| i \left>\right< i \right| \left| j \left>\right< j \right|   \otimes \mathcal{S}(\N{1},\ldots,{\mathcal N}_{N} ) (\rho' \otimes \rho_c) \right] \nonumber \\
	& = \frac{\mathbb{I}}{d} \otimes {\tilde \rho}_c^{(N)} 
\end{align}
\noindent here $\rho'=X(i) Z(j) \rho Z(j)^\dagger X(i)^\dagger$ and $X(i)\left| l \right>=\left|i \oplus l \right>, Z(j)\left| l \right>=e^{2 \pi i j l} \left| l \right>$ are the known Heisenberg$-$Weyl operators \cite{wilde2013quantum}. To isolate the term ${\tilde \rho}_c^{(N)} $ we apply the following relations 
\begin{eqnarray} \label{rules2}
	{\rm Tr}_{XIJ} \left[ \sum_{XIJ} p_x \left| x \left>\right< x \right| \left| i \left>\right< i \right| \left| j \left>\right< j \right| \rho' \right] = d \mathbb{I}  \\
	{\rm Tr}_{XIJ} \left[ \sum_{XIJ} p_x \left| x \left>\right< x \right| \left| i \left>\right< i \right| \left| j \left>\right< j \right| \mathbb{I} \right] = d^2 \mathbb{I} 
\end{eqnarray}
which are valid for $N\geq2$ and they are obtained by direct calculation following the former definitions.
%
%
Then, for $N=2$ we find that the output control state is
\begin{equation}  \label{output2}
	{\tilde \rho}_c^{(2)} = p_1 p_2 [ P_1 \left| 1 \left>\right< 1 \right| + P_2 \left| 2 \left>\right< 2 \right| +
	%
	\frac{\sqrt{P_1P_2}}{d^2} \left( \left| 0 \left>\right< 1 \right| + \left| 1 \left>\right< 0 \right| \right) ] + \rho_c \left( 1 - p_1 p_2 \right)
\end{equation}
\noindent where $p_i=1-q_i$.  

\begin{figure} [h!]
	\vspace*{13pt}
		\scalebox{.1}{\includegraphics{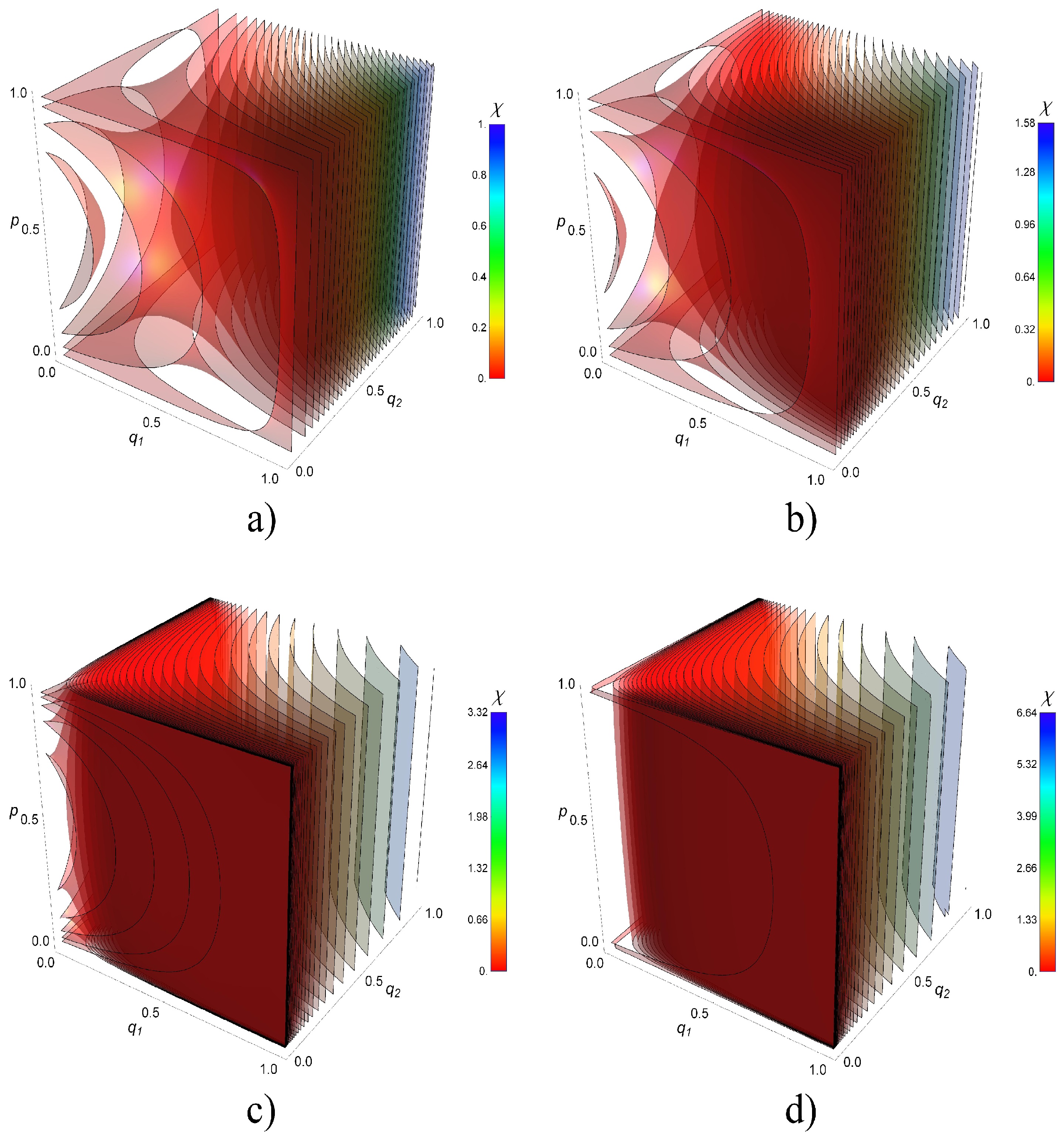}}
	\vspace*{13pt}
	\caption{\label{Holevo2} {Transmission map of information for two noisy channels.}The 3D graphs represent contour surfaces of the Holevo information $\chi_{\rm Q2S}$ when the depolarising parameters $q_1,q_2$ and the probabilities $P_1=P_2=p$ varied from 0 to 1. We plot several cases for the dimension $d$ of the target system: a) $d=2$, b) $d=3$, c) $d=10$ and d) $d=100$. In all these cases there are thirty contour surfaces of $\chi_{\rm Q2S}$. The values of $\chi_{\rm Q2S}$ are shown in the color bars. } 
\end{figure}

Using the two previous results for $\QNS$, Figure \ref{Holevo2} shows the transmission map  of information for two noisy channels. The plots are contour surfaces of $\chi_{\rm Q2S}$ when $q_1,q_2$ and $P_1=P_2=p$ vary from 0 to 1. The maximum capacity is trivially reached when $q_1=q_2=1$ simultaneously reaching the value $\chi_{\rm Q2S}=\log d$.
The minimum capacity is zero, reached in the boundary of the front sides with $(q_1=0, p=0,1)$, $(q_2=0, p=0, 1)$, $(q_1=0, q_2=1)$, and $(q_1=1, q_2=0)$. Notably, for $q_1=q_2=0$ there are values higher than the minimum. This phenomenon is  observed in the protuberance of plots near $\chi_{\rm Q2S}=0$. 
For larger values of $d$, the protuberance occurs sharply near  $q_1=0$ and $q_2=0$ faces. Note the nearest surface to those faces are for $\chi_{\rm Q2S}=10^{-3}, 10^{-3}, 10^{-4}, 10^{-7}$ respectively for each plot $d=2, 3, 10, 100$. 
%

\subsection{Holevo information  for $N=3$ channels}\label{CalculationHmin3channels}

We numerically calculate the eigenvalues of the entropy $H^{\rm min}$ for $N=3$ channels from equation (\ref{Matrix3i}). Then, using relations (\ref{rules2}) from (\ref{S0}), (\ref{S1}), (\ref{S2}) and (\ref{S3}) we find that the output state is

\begin{eqnarray}\label{output3}
	{\widetilde \rho}_c ^{(3)}= (s_2+s_3)  \rho_c +\displaystyle \frac{s_0}{d^2} \left(\sum_{(k,k')\in {\mathfrak I},{\mathfrak J}} \sqrt{P_k P_{k'}} \vert k\rangle\langle k'\vert \right.
	%
	+ \displaystyle \left. d^2 \sum_{(k,k')\in {\mathfrak K}} \sqrt{P_k P_{k'}} \vert k\rangle \langle k'\vert \right) \\
	\hspace{1.5cm} +\displaystyle \frac1{d^2} \left(\sum_{s=1}^3 \sum_{(\ell,\ell')\in {\mathfrak L}_s} \sqrt{P_\ell P_{\ell'}} r_s \vert \ell \rangle \langle \ell'\vert
	%
	+ d^2 \displaystyle \sum_{s=1}^3 \sum_{(m,m')\in {\mathfrak M}_s} \sqrt{P_m P_{m'}} r_s \vert m \rangle \langle m'\vert \right) \nonumber
	%
\end{eqnarray}

\begin{figure} [h!]
	\vspace*{13pt}
\scalebox{.6}{\includegraphics[width = .7\textwidth]{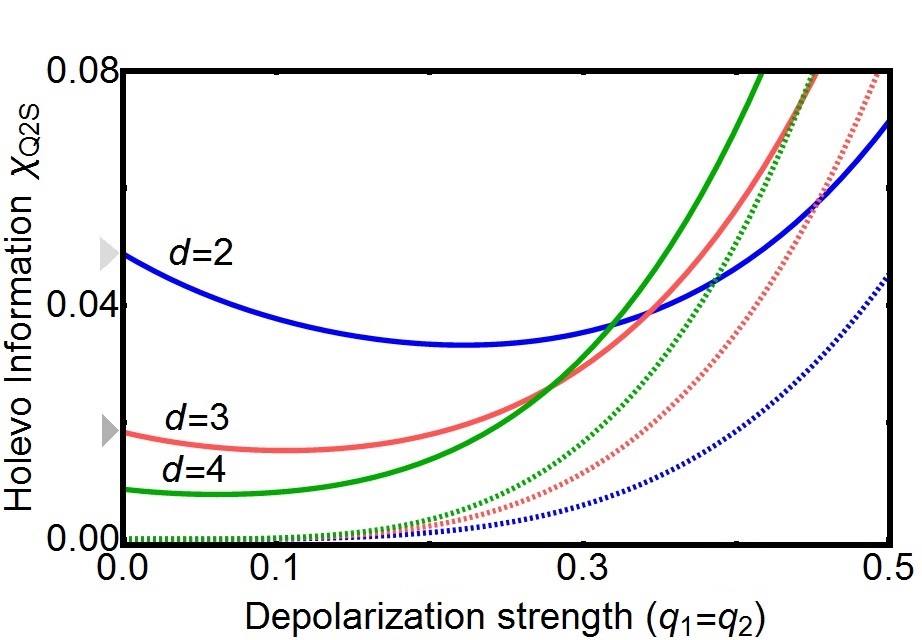}}(a)\\
\scalebox{.6}{\includegraphics[width = .7\textwidth]{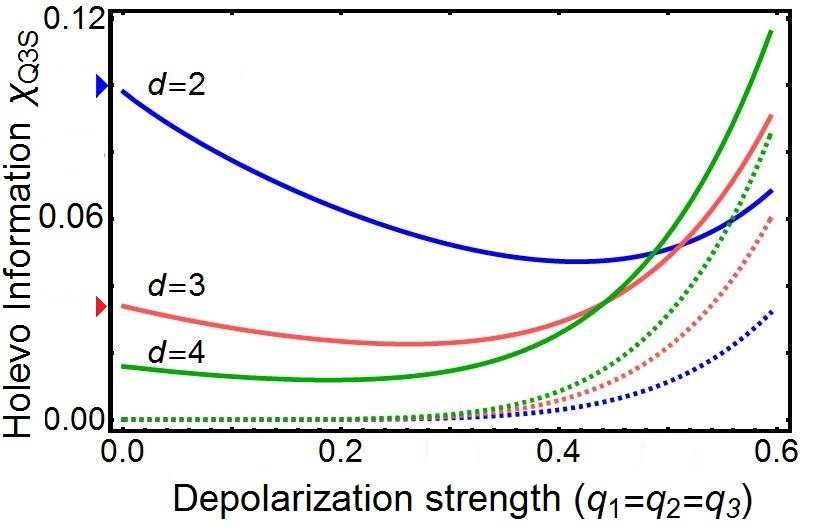}}(b)
	\vspace*{13pt}
	\caption{\label{Figura3}{Transmission of information for $N=2$ and $N=3$ channels.} Holevo information as a function of the depolarization strengths $q_i$ of the channels. We plot the subcases of equal depolarization strengths, i.e., $q_1=q_2=q_3=q$, with  equally weighted probabilities $P_k$ for indefinite causal orders (solid line) with (a) $N=2$ and (b) $N=3$ channels.  The transmission of information first decreases to a minimal value for Holevo information and then the transmission of information increases with $q$. For completely depolarizing channels, i.e. $q=0$, the transmission of information is nonzero and decreases as $d$ increases. A comparison is shown between the Holevo information when the channels are in a definite causal order (dashed line). A full superposition of $N!$ causal orders is used.} 
\end{figure}

As before, putting those outcomes together in (\ref{Gholevo}), Figures~\ref{Figura3} (a) and (b) give the Holevo informations $\chi_{\text{Q2S}}$ and $\chi_{\text{Q3S}}$ for two and three channels respectively, as a function of the depolarization strengths $q_i$  and the dimension $d$ of the target system.  Our model enables us to exhibit a wealth of different behaviors as a function of $d$,$q_i$, and $N$ from the fully noisy situation to the identity channel transmission. For the sake of simplicity, we restrict our graphical analysis to equal depolarization strengths, i.e., $q_1=q_2=q_3$, with a balanced superposition of $N!$ causal orders, that is, with equally weighted probabilities $P_k=1/N!$ for each case $N=2,3$. 
The analysis of these results allows us to draw the following conclusions for those particular cases: 
\begin{itemize} 
	\item For a fixed dimension $d$, the Holevo information for indefinite causal order is always higher than the one obtained using one of the definite causal order shown in Fig.~\ref{Figura2}. This is especially the case  for totally depolarized channels i.e. $q_i=0, \forall i$. For completely clean channels ($q=1$), the Holevo information for indefinite and definite causal order converges to the same value depending on $d$ (not shown).
	\item Two regions can be distinguished. In the strongly depolarized region (roughly $q<0.3$ for $N=2$ and $q<0.5$ for $N=3$) the increase of the dimension $d$ of the target system is detrimental to the Holevo information transmitted by the quantum switch. In contrast, in the moderately depolarized region ($q>0.3$ for $N=2$ and $q>0.5$ for $N=3$) the Holevo information increases both with $q$ and $d$,  as expected a maximum (not shown) for completely clean channels. 
	\item In the strongly depolarized region, increasing the number of channels to $N=3$ is definitively advantageous for information extraction. For instance, in the case of totally depolarized channels ($q=0$), the Holevo information is approximately doubled with $N=3$ with respect to $N=2$ for all values of the dimension $d$ calculated up to $d=10$
\end{itemize}

In fact, Table \ref{table1} gives the values of the ratio $\chi_{\text Q3S}/\chi_{\text Q2S}$, finding that the Holevo information is approximately doubled for $N=3$ with respect to $N=2$.

\begin{table}[ht]
	\centering 
	\begin{tabular}{c c c c} 
		\hline 
		$d$ & $\chi_{\text Q2S}$  & $\chi_{\text Q3S}$  & $\chi_{\text Q2S}/\chi_{\text Q3S}$ \\ [0.5ex] 
		\hline 
		2 & 0.0487 &0.0980 & 2.0123 \\ 
		3 & 0.0183 & 0.0339 & 1.8524 \\
		4 & 0.0085 & 0.0159 &1.8705 \\
		5 & 0.0046 & 0.0087 & 1.8913 \\
		6 &0.0027 & 0.0053 & 1.9629 \\
		7 &0.0018 & 0.0034 & 1.8888 \\
		8 & 0.0012 & 0.0023 & 1.9166 \\
		9 & 0.0008 & 0.0016 & 2 \\
		10 & 0.0006 & 0.0012 & 2 \\ [1ex] 
		\hline 
	\end{tabular}
	\caption{Values of the Holevo information ratio $\chi_{\text Q3S}/\chi_{\text Q2S}$. The mean value of the ratio is 1.9328 $\pm$ 0.0617.} 
	\label{table1} 
\end{table}

\section{Conclusions}\label{concl}
Communication enhancement is a challenging task in quantum information processing due to imperfection of communication channels  subjected to depolarization. Causal order has been proposed as a disruptive procedure to improve communication, compression of quantum information, bringing the quantum possibilities into a new frontier. We have analyzed the quantum control of $N$ operators in the context of the second-quantized Shannon theory and in the specific case of superposition of causal orders,  extending the results in the current literature. We obtained a general expression for ${\mathcal S}(\N{1},\N{2},\ldots,\N{N})$for the quantum $N$-switch for an arbitrary number of channels with any depolarizing strength thus providing an operational formula enabling the exploration of communication channels controlled by causal orders. This formula is useful to explore computationally the cases with an increasing $N$. A detailed analysis to assess the information transmission for the cases of $N=2$ and $N=3$ channels is presented:  an increasing number of  channels improves the  transmission of information. In particular, we remarkably found that the Holevo information  is doubled when the number of channels goes from $N=2$ to $N=3$.

We give the matrices  corresponding to quantum $N$-switches ${\mathcal S}(\N{1},\N{2},\ldots,\N{N})$ as a function of the number of channels, depolarization strengths, dimension of the target system. We   obtain other general properties for the general case of $N$ channels such as the symmetric properties of matrices $Q^{k,k'}_{A_z}$ and thus of ${\mathcal S}(\N{1},\N{2},\ldots,\N{N})$. We also demonstrate that ${\mathcal S}(\N{1},\N{2},\ldots,\N{N})$ is always a linear combination of $\rho$ and $\mathbb{I}_t$, whatever $N$ channels. Expressions for the Holevo limit are equally accessible from our expressions and methodology. Besides, we showed that the depolarizing strengths can be used as control parameters to modify the information transmission on demand. We shall develop elsewhere the analysis of control via selected combinations of the $N!$ available causal order enabled by the present work.
\section*{Acknowledgements}
\noindent
F. Delgado acknowledges professor Jes\'us Ram\'irez-Joach\'in for the fruitful discussions and teaching in 1983 about combinatorics required in this work. L.M. Procopio wishes to thank   Alastair A. Abbott for commenting this manuscript. F. Delgado and M. Enr\'iquez acknowledge the support from CONACyT and from School of Engineering and Science of Tecnol\'ogico de Monterrey in the developing of this research work. L.M. Procopio acknowledges the support from the European Union's Horizon 2020 research and innovation
programme under the Marie Sk\l{}ukodowska-Curie grant agreement No 800306.  This work is also supported by a public grant overseen by the French National Research Agency (ANR) as part of the ``Investissements d'Avenir'' program (Labex NanoSaclay, reference: ANR-10-LABX-0035) and by the Sitqom ANR project  (reference : ANR-SITQOM-15-CE24-0005).\\


\appendix 

\section{Completeness property for $W_{\mathbf{i}}$} \label{AppA}
\numberwithin{equation}{section}
\setcounter{equation}{0}

We demonstrate here the completeness property 
\begin{equation}\label{Wproperty}                                                   
	\sum_{\{i_s\}|_{s=1}^N}  W_{\mathbf{i}} W_{\mathbf{i}}^\dagger= \mathbb{I}_t \otimes \mathbb{I}_c,
\end{equation}
\noindent for the generalized Kraus operators $W_{\mathbf{i}}$ for the full quantum $N$-switch channel by relying on the reordering of the sums obtained by grouping terms with indices $i_s$ equal to zero. $W_{\mathbf{i}} := W_{i_1 i_2\ldots i_N}=  \sum_{k=1}^{N!} K_{\pi_k} \otimes \left| k \right>\left< k \right|$ and  	$K_{\pi_k} := \pi_k( K_{i_1}^{1}\cdots K_{i_N}^{N})$ where $\pi_k$ acts on the subscripts $j$ of the Kraus operators $ K_{i_j}^{j}$. In the sum $\{i_j\}|_{j=1}^N$, each index in the set of indices $\{i_1,i_2,\ldots,i_N\}$  is associated to a channel $\N{j}$ where $j\in \{1,2,\ldots,N\}$ and varies from $0$ to $d^2$.  
By introducing the definition of the Kraus operators  $ K_{i_j}^{j} = \frac{\sqrt{1-q_j}}{d} U_{i_j}^{j}$ into $W_{{\mathbf{i}}}$, the left side from Equation~(\ref{Wproperty}) can be re-written as  
\begin{equation} 
	h_Nd^{-2N} \sum_{\{i_s\}|_{s=1}^N} \sum_{k=1}^{N!} U_{\pi_k} U_{\pi_k}^\dagger \otimes \left| k \right> \left< k \right|,
\end{equation}
where  
$U_{\pi_k} := \pi_k( U_{i_1}^{1}\cdots U_{i_N}^{N})$  and $h_N=\prod_{j=1}^{N} 1-q_j$.
As $U_i U_i^\dagger=\mathbb{I}$, $\forall i> 0$, the product $ U_{\pi_k} U_{\pi_k}^\dagger $ reduces almost to the identity $\mathbb{I}_t$ except for the factors $U_0U_0^\dagger=\frac{d^2q_j}{1-q_j}\mathbb{I}_t$. Using (\ref{sumABz}) in the above expression, it can be rearranged into:
\begin{equation}
	\sum_{\{i_j\}|_{j=1}^N} W_{i_1 i_2\ldots i_N} W^\dagger_{i_1 i_2\ldots i_N} =  h_Nd^{-2N}\sum_{k=1}^{N!} \sum_{z=0}^N d^{2z}  \sum_{a\in {\bf A}_z^N}  \sum_{b\in B_z} h_{A_z} \mathbb{I}_t \otimes \left| k \right>\left< k \right|, 
\end{equation}
where the sum over $A_z$ is the sum of terms $h_{A_z}$ over all the elements of ${\bf A}_z^N$, the set of all subset of $z$ elements in $\{1,2,\ldots,N\}$. This yields the factor $f_{A_z}=h_Nh_{A_z}d^{2(z-N)}$ in equation~(\ref{Sz}). 

To prove equation~(\ref{Wproperty}), we then apply the total probability property $ \sum_{z=0}^N \sum_{a\in A_z} h_{A_z} = \frac{1}{h_N}$ together with the property $\sum_{b\in B_z}d^{2(z-N)}=1$. Thus (\ref{Wproperty}) is then proved.


\section{Relations to evaluate coefficients $Q^{k,k'}_{A_z}$} \label{Rules}

We recall below the relations needed to deduce explicitly matrices $\mathcal{S}_z$ and then $\mathcal{S}$ for the quantum $N$-switch from the sums and products of the $Q^{k,k'}_{A_z}$ factors

\begin{gather}
	\label{unitaryproperty1}
	\sum_{i=1}^{d^2} U_i^{}X\left[U_i^{} \right]^\dagger = d {\tr} X ~ \mathbb{I}\\
	\label{unitaryproperty2}
	\sum_{i=1}^{d^2} {\tr}([U_i^{}]^\dagger \rho) U_i^{}= \sum_{i=1}^{d^2} {\tr}(U_i^{} \rho) \left[U_i^{} \right]^\dagger=d~ \rho
\end{gather}
\noindent where $X$ is any $d \times d$ matrix and $U_i$ an orthonormal basis for the $d\times d$ matrices. Applying equation (\ref{unitaryproperty1}) to $X=\mathbb{I}$,
\begin{equation}
	\label{unitaryproperty3}
	\sum_{i=1}^{d^2} U_i^{}\left[U_i^{} \right]^\dagger = d^2 \mathbb{I}.
\end{equation}
\noindent Applying equation (\ref{unitaryproperty1}) to $X=\rho$, such that ${\rm Tr}(\rho)=1$, we get a uniform randomization over the set of unitaries $U_{i\neq 0}$ that completely depolarizes the state $\rho$, thus giving the relation $ \sum_{i} U_{i} \rho U_{i}^{\dagger}=d \mathbb{I}$.

\section{Matrices $\mathcal{S}_z$ for the quantum 3-switch} \label{AppC}
\noindent Matrix ${\mathcal S}(\N{1},\N{2},\N{3})(\rho\otimes\rho_c)$  is a $6\times 6$ block-matrix, block-symmetric matrix whose matrix elements are matrices of dimension $d\times d$. Since  $Q^{k,k'}_{A_z} = Q^{k',k}_{A_z}$, for all $A_z$, then  ${\mathcal S}(\N{1},\N{2},\N{3})(\rho\otimes\rho_c)$ is symmetric with respect to the main diagonal. Now, by expanding equations $\mathcal{S}_0$, $\mathcal{S}_1$, $\mathcal{S}_2$ and $\mathcal{S}_3$ in the control qudit basis, $\{\left|1\right>,  \left|2\right> ,\left|3\right>,\left| 4\right>,\left|5\right>,\left|6\right>\}$, we can  found the quantum 3-switch  matrix ${\mathcal S}(\N{1},\N{2},\N{3})(\rho\otimes\rho_c)$ having diagonal elements as 
\begin{equation}
	\mathcal{A}_k=P_k[(s_0+s_2+t_1+t_2+t_3) \mathbb{I}/d+s_3\rho], \quad \quad \mbox{ for } k=1,2,\ldots,6,
\end{equation} 
and off-diagonal elements
\begin{equation}
	\begin{array}{ll}
		\mathcal{B}=\sqrt{P_1P_2}(d^2 s_2+d^2 t_2+d^2 t_3+s_0)\mathbb{I}/d^3
		+\sqrt{P_1P_2}(d^2 s_3+t_1)\rho/d^2,\\[1em]%
		\mathcal{C}=\sqrt{P_1P_3}(d^2 s_2+d^2 t_1+d^2 t_2+s_0)\mathbb{I}/d^3
		+\sqrt{P_1P_3}(d^2 s_3+t_3)\rho/d^2,\\[1em]%
		\mathcal{D}=\sqrt{P_1P_4}(t_1+s_2)\mathbb{I}/d
		+\sqrt{P_1P_4}(d^2 s_3+s_0+t_2+t_3)\rho/d^2,\\[1em]%
		\mathcal{E}=\sqrt{P_1P_5}( s_2+t_3)\mathbb{I}/d
		+\sqrt{P_1P_5}(d^2 s_3+s_0+t_1+t_2)\rho/d^2,\\[1em]%
		\mathcal{F}=\sqrt{P_1P_6}(d^2 s_2+s_0)\mathbb{I}/d^3
		+\sqrt{P_1P_6}(d^2 s_3+t_1+t_2+t_3)\rho/d^2,\\[1em]%
		\mathcal{G}= \sqrt{P_2P_3}(s_2+t_2)\mathbb{I}/d
		+\sqrt{P_2P_3}(d^2 s_3+s_0+t_1+t_3)\rho/d^2,\\[1em]%
		\mathcal{H}= \sqrt{P_2P_4}(d^2 s_2+s_0)\mathbb{I}/d^3
		+\sqrt{P_2P_4}(d^2 s_3+t_1+t_2+t_3)\rho/d^2,\\[1em]%
		\mathcal{I}= \sqrt{P_2P_5}(d^2 s_2+d^2 t_1+d^2 t_3+s_0)\mathbb{I}/d^3
		+\sqrt{P_2P_5}(d^2 s_3+ t_2)\rho/d^2, \\[1em]%
		\mathcal{J}= \sqrt{P_2P_6}(s_2+t_1)\mathbb{I}/d
		+\sqrt{P_2P_6}(d^2 s_3+s_0+t_2+t_3)\rho/d^2, \\[1em]%
		\mathcal{K}= \sqrt{P_3P_4}(d^2 s_2+d^2 t_1+d^2 t_3+s_0)\mathbb{I}/d^3
		+\sqrt{P_3P_4}(d^2 s_3+ t_2)\rho/d^2, \\[1em]%
		\mathcal{L}= \sqrt{P_3P_5}(d^2 s_2+s_0)\mathbb{I}/d^3
		+\sqrt{P_3P_5}(d^2 s_3+t_1+t_2+t_3) \rho/d^2, \\[1em]%
		\mathcal{M}= \sqrt{P_3P_6}(s_2+t_3)\mathbb{I}/d
		+\sqrt{P_3P_6}(d^2 s_3+s_0+t_1+t_2) \rho/d^2, \\[1em]%
		\mathcal{N}= \sqrt{P_4P_5}(s_2+t_2) \mathbb{I}/d
		+\sqrt{P_4P_5}(d^2 s_3+s_0+t_1+t_3) \rho/d^2, \\[1em]%
		\mathcal{P}= \sqrt{P_4P_6}(d^2 s_2+d^2 t_2+d^2 t_3+s_0) \mathbb{I}/d^3
		+\sqrt{P_4P_6}(d^2 s_3+ t_1) \rho/d^2, \\[1em]%
		\mathcal{Q}= \sqrt{P_5P_6}(d^2 s_2+d^2 t_1+d^2 t_2+s_0) \mathbb{I}/d^3
		+\sqrt{P_5P_6}(d^2 s_3+t_3) \rho/d^2.%
	\end{array}
\end{equation}
\noindent 
Those matrix elements are the entries of the matrix~(\ref{Matrix3i}) in the main text.

\end{document}